\documentclass{aa}  

\usepackage{graphicx}	
\usepackage[dvipsnames]{xcolor}
\usepackage[varg]{txfonts}
\usepackage{afterpage}
\usepackage{amsmath}	

\bibpunct{(}{)}{;}{a}{}{,} 

\usepackage{xpatch}
\usepackage{hyperref}
\hypersetup{
	colorlinks=true,
	breaklinks=true,
	citecolor=blue,
	allcolors=blue,
	frenchlinks=true
}

\makeatletter
\xpatchcmd\NAT@citex
{%
	\@citea\NAT@hyper@{%
		\NAT@nmfmt{\NAT@nm}%
		\hyper@natlinkbreak{\NAT@aysep\NAT@spacechar}{\@citeb\@extra@b@citeb}%
		\NAT@date
	}%
}
{%
	\@citea
	\NAT@nmfmt{\NAT@nm}%
	\NAT@aysep\NAT@spacechar
	\NAT@hyper@{\NAT@date}%
}
{}{}
\xpatchcmd\NAT@citex
{%
	\@citea\NAT@hyper@{%
		\NAT@nmfmt{\NAT@nm}%
		\hyper@natlinkbreak{\NAT@spacechar\NAT@@open\if*#1*\else#1\NAT@spacechar\fi}%
		{\@citeb\@extra@b@citeb}%
		\NAT@date
	}%
}
{
	\@citea
	\NAT@nmfmt{\NAT@nm}%
	\NAT@spacechar\NAT@@open\if*#1*\else#1\NAT@spacechar\fi
	\NAT@hyper@{\NAT@date}%
}
{}{}
\makeatother

\newcommand{\bs}[1]{\boldsymbol{#1}}
\newcommand{\bcdot}{\boldsymbol{\cdot}}
\newcommand{\bnabla}{\boldsymbol{\nabla}}

\begin{document} 

   \title{CRexit observed: probing cosmic ray transport in the circumgalactic medium with absorption line spectra}
   \titlerunning{CRexit observed: probing cosmic ray transport in the CGM}

   \author{M. Weber\inst{1}
          \and
          T. Thomas\inst{1}
          \and
          C. Pfrommer\inst{1}
          \and
          T. Urrutia\inst{1}
          }

    \institute{
    Leibniz institute for astrophysics in Potsdam (AIP), An der Sternwarte 16, D-14482 Potsdam, Germany\\
    \email{maweber@aip.de}
    }

   \date{\today}

\abstract{
Cosmic rays (CRs) likely provide dynamically important non-thermal pressure support in the circumgalactic medium (CGM), but how their transport physics shapes observable absorption signatures remains uncertain. We investigate whether absorption-line diagnostics can distinguish between different CR transport regimes in CR-pressure-dominated halos. Using high-resolution simulations, we generate synthetic spectra along large ensembles of sightlines and measure column densities, equivalent widths, covering fractions (CFs), velocity widths, abundance ratios, and stacked absorption profiles for ions tracing cool, warm, and hot gas. We find that the effective CR transport speed strongly regulates the multiphase structure of the CGM. Efficient CR transport enhances the formation of cool ($T\sim10^4\,\mathrm{K}$) and warm ($T\sim10^5\,\mathrm{K}$) gas, leading to deeper and broader absorption lines of low- and intermediate-ionization species. The two-moment CR transport model produces the strongest \ion{Mg}{II} and \ion{Si}{II} absorption and reaches \ion{Mg}{II} CFs consistent with the range inferred for star-forming galaxies. In contrast, slow CR transport underproduces cool, low-ionization gas and yields substantially weaker absorption. We also find that the origin of \ion{C}{IV}-bearing gas changes with CR transport: slow transport mainly produces extended warm halo gas, whereas efficient transport shifts much of the \ion{C}{IV} absorption into mixing layers around cool clouds. The high-ionization tracer \ion{O}{VI} responds more weakly, indicating that CR transport primarily regulates the cool condensed phase and its interfaces rather than the volume-filling hot halo. These findings suggest that absorption-line measurements of cool and transition-phase gas can provide valuable constraints on the effective transport of CRs through the CGM.
}


\maketitle

\nolinenumbers
%

\section{Introduction}

The circumgalactic medium (CGM) is a gaseous reservoir between galaxies and their large-scale, cosmological environment that regulates the exchange of gas, metals, and energy, thereby playing an important role in the galactic baryon cycle \citep{Peroux2020}. Rather than being a smooth and diffuse atmosphere, the CGM is highly structured and multiphase, with cool ($T\sim10^4\,\mathrm{K}$), warm ($T\sim10^5\,\mathrm{K}$), and hot ($T\gtrsim10^6\,\mathrm{K}$) gas coexisting over a wide range of densities and galactocentric radii \citep{Faucher-Giguere2023}. This complex structure is most directly probed by absorption-line spectroscopy of background sources, which reveals cool low-ionization gas coexisting with warm-hot, highly ionized material far beyond the stellar disk \citep{Chen2026}. Since a substantial fraction of a galaxy's baryons and metals resides in this extended reservoir, the CGM plays a central role in connecting feedback, gas accretion, and the long-term evolution of galaxies \citep{Tumlinson2017}. Understanding the physical processes that establish and maintain its multiphase structure is therefore a key challenge for models of galaxy formation.

A central theoretical challenge is to understand how the cool CGM gas observed around galaxies forms and survives within a much hotter halo. At low redshift, cosmological accretion is expected to proceed predominantly through hot-mode inflows, limiting the direct supply of cool material from the intergalactic medium \citep{Keres2005, Nelson2013, Afruni2023, Stern2024}. Although galactic outflows can contribute to the cool CGM through recycling, entrainment, and cooling within winds \citep{Fielding2022, Gronke2022}, directly launched cool clouds alone appear insufficient to account for the full observed low-redshift cool-gas population without invoking extreme outflow energetics \citep{Afruni2021, Decataldo2024}. This has motivated scenarios in which cool gas forms in situ through thermal instability (TI) and precipitation within the hot CGM \citep{Field1965}. In this picture, local overdensities condense when radiative cooling becomes efficient compared to dynamical timescales, producing cool clouds embedded in the diffuse halo \citep{Sharma2012, McCourt2012, Voit2017}. However, both the efficiency of condensation and the survival of the resulting clouds depend sensitively on the additional physics that shapes the CGM, including halo mass, turbulence, magnetic fields, feedback, and non-thermal pressure support \citep{Fielding2017a, Lochhaas2020, Augustin2025, Mohapatra2025, Thomas2025a, Gronke2026}.

Cosmic rays (CRs) provide a natural mechanism for modifying the thermal and dynamical state of the CGM. As a non-thermal component with energy densities comparable to those of the thermal gas, magnetic fields, and radiation, CRs can contribute significantly to the pressure balance in galactic environments \citep{Zweibel2013, Grenier2015, Ruszkowski2023}. Their relevance for the CGM is supported by both observations and numerical models. Observed extended radio-continuum halos around disk galaxies trace synchrotron emission from CR electrons and magnetic fields far above the stellar disc, indicating that non-thermal components are transported efficiently into galactic halos \citep{Wiegert2015, Heesen2021, Stein2023}. Numerical studies and synthetic non-thermal-emission models further support this picture, showing that CRs can propagate into the CGM and leave observable radio signatures on halo scales \citep[e.g.][]{Werhahn2021c, Pfrommer2022, Chiu2024, Ponnada2024}. In this regime, CR pressure can rival or even exceed the thermal pressure, producing partially or fully CR-pressure-dominated halos, as seen in high-resolution interstellar medium tall-box setups \citep[e.g.][]{Girichidis2016, Girichidis2018, Simpson2016, Bustard2020, Sike2025, Kim2026}, isolated galaxy simulations \citep[e.g.][]{Uhlig2012, Booth2013, Salem2016, Pakmor2016c, Pfrommer2017b, Butsky2022, Farcy2022, Thomas2023, Thomas2025b, Kjellgren2025}, and cosmological zoom-in simulations of galaxy formation \citep[e.g.][]{Buck2020, Ji2020, Hopkins2020, Hopkins2021a, Bieri2026}. Such non-thermal pressure support has the potential to strongly affect the formation and survival of multiphase gas. By altering the force balance, buoyancy, and compressibility of the halo gas, CRs can suppress the growth of overdensities and regulate whether gas condenses or remains supported by hot pressure. This makes CRs a promising ingredient for explaining the extended cool-gas distributions seen around star-forming galaxies, especially since purely thermal feedback models often struggle to reproduce galaxy properties and CGM observables simultaneously \citep[e.g.][]{Butsky2018, Chan2019, DeFelippis2024, Lu2026}.

The extent to which CRs reshape the CGM, however, depends not only on their pressure contribution, but also on how efficiently they are transported relative to the thermal gas \citep[e.g.][]{Thomas2020}. In CR-pressure-dominated halos, tightly coupled CRs with negligible or very slow transport relative to the gas are compressed together with cooling overdensities and can build up additional non-thermal pressure support. This stabilizes dense gas against further collapse and can suppress the formation of cold clouds. If, instead, CRs stream or diffuse efficiently along magnetic field lines, they can escape from overdense regions and reduce the local CR pressure support, allowing the gas to cool and condense more readily \citep{Sharma2010, Butsky2020, Ji2020, Weber2025}. CR transport therefore determines whether CRs primarily stabilize the halo gas or promote the formation of a multiphase medium. Despite its central importance, the effective transport of CRs in the weakly collisional CGM remains uncertain. CRs may stream along magnetic field lines, diffuse anisotropically with an energy-dependent coefficient \citep{Girichidis2022}, or interact with the plasma through more complex wave-particle processes \citep{Armillotta2022, Kempski2023, Reichherzer2025, Sike2025, Thomas2025b, Ewart2026}. This uncertainty makes it essential to identify observable signatures that can distinguish between different CR transport regimes.

\begin{figure}[!ht]
    \centering
    \includegraphics[width=\linewidth]{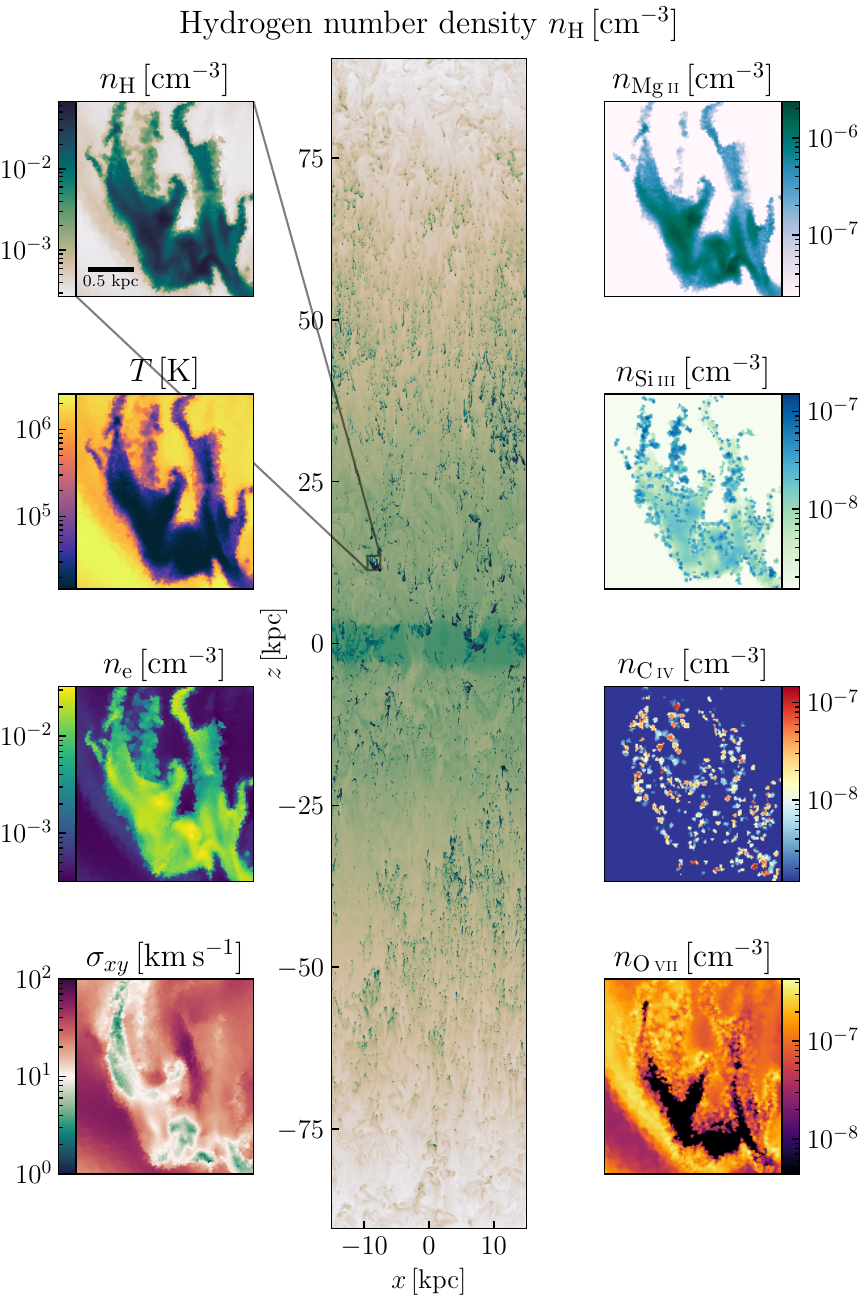}
    \caption{Thin projections through the simulation domain. The central panel shows the hydrogen number density, $n_\mathrm{H}$, across the full $30\times180\,\mathrm{kpc}$ extent in the $x$--$z$ plane, highlighting the filamentary and clumpy structure of the multiphase CGM. The panels on the left display a zoom-in on a representative cold cloud, showing $n_\mathrm{H}$, temperature $T$, electron number density $n_\mathrm{e}$, and turbulent velocity dispersion $\sigma_{xy}$.
    The panels on the right present the corresponding ion number densities of \ion{Mg}{II}, \ion{Si}{III}, \ion{C}{IV}, and \ion{O}{VII}, demonstrating how low-ionization species trace the dense, cold core, intermediate-ionization species trace mixing layers at cloud boundaries, and high-ionization species arise predominantly in the surrounding hot ambient medium.
    }
    \label{fig:singleCloudZoom}
\end{figure}

If CR transport regulates the phase structure of the CGM, its effects should be imprinted on absorption-line observables. Efficient transport is expected to change not only the amount of cool and warm gas, but also its spatial covering fraction (CF) and line-of-sight velocity structure. These changes can be tested with ionic tracers that sample different gas phases: low-ionization species such as \ion{Mg}{II} and \ion{Si}{II} trace cool, condensed gas, intermediate ions such as \ion{C}{IV} probe warmer gas and cloud--halo interfaces, and high ions such as \ion{O}{VI} trace the more diffuse warm-hot halo \citep{Rauch1998, Quider2011, Tumlinson2011, Werk2016}. Observed CGM absorption systems show large CFs, a broad range of equivalent widths (EWs), and complex velocity structure \citep[e.g.][]{Nielsen2013a, Kacprzak2015, Bordoloi2017, Chen2017}, indicating that the halo gas is multiphase and dynamically active. However, the physical origin of these signatures remains uncertain. In particular, it is unclear whether the observed quantities can be reproduced by thermal condensation alone, or whether non-thermal components such as CRs are required. Forward-modeling these observables from simulations with different CR transport prescriptions therefore provides a direct way to connect CR microphysics to observable CGM diagnostics.

In this work, we investigate how different prescriptions for CR transport shape the observable signatures of CR-pressure-dominated CGMs. Using high-resolution simulations, we generate synthetic spectra along a large ensemble of sightlines and measure column densities, EWs, CFs, velocity widths, and abundance ratios for key ionic tracers spanning multiple gas phases. By systematically comparing these diagnostics across models with different CR transport efficiencies, we aim to identify robust absorption-line signatures that can constrain the role of CRs in regulating the CGM. This paper is organized as follows. In Sec.~\ref{sec:setup}, we describe the numerical setup and the construction of curved sightlines used for our mock absorption spectra. In Sec.~\ref{sec:theoreticalbackground}, we briefly summarize the observational diagnostics employed in this work. The resulting column-density distributions, absorption-line profiles, EW statistics, CFs, abundance ratios, and velocity distributions are presented in Sec.~\ref{sec:results}. In Sec.~\ref{sec:discussion}, we discuss the limitations of our model and summarize our conclusions. Details of the ray-tracing algorithm and the absorption-line synthesis are provided in Appendices~\ref{app:raytracing} and \ref{app:absorptionlinesynthesis}, respectively. Appendix~\ref{app:mgII_doublet_ratio} presents an analysis of the \ion{Mg}{II} $\lambda\lambda2796,2803$ doublet ratio and the resulting saturation of the \ion{Mg}{II} absorption, while Appendix~\ref{app:path_length_effect} quantifies the geometric contribution to the radial trends in our synthetic observables.

\section{Numerical model and synthetic observations}
\label{sec:setup}
For the present study, we analyze the simulations introduced in \citet{Weber2025} and post-process them to generate synthetic absorption-line spectra. In the following, we briefly summarize the relevant aspects of the numerical setup and describe the modeling procedure used to connect the simulations to observable CGM diagnostics.

\subsection{Simulation framework}

We performed three-dimensional CR magnetohydrodynamic (CRMHD) simulations with the moving-mesh code \textsc{Arepo} \citep{Springel2010, Pakmor2016}. The coupled MHD and CR fluid equations are solved with a finite-volume scheme on an unstructured Voronoi mesh. CR transport is modeled using a two-moment method that evolves both the CR energy density and flux, thereby capturing streaming and diffusive transport within a unified framework \citep{Thomas2019, Thomas2021}. The mesh is adaptively refined and de-refined to maintain an approximately constant gas mass per cell. This increases the spatial resolution in dense, cooling structures while preserving computational efficiency in the diffuse halo gas. The computational domain represents a stratified vertical column of the CGM, extending to $\pm 90\,\mathrm{kpc}$ from the galactic midplane, with a total box size of approximately $30 \times 30 \times 180\,\mathrm{kpc}$ (cf. Fig.~\ref{fig:singleCloudZoom}). The target gas mass resolution is $m_\mathrm{target}=318\,\mathrm{M}_\odot$. We imposed periodic boundary conditions in the horizontal directions and open boundaries in the vertical direction to allow gas to leave the domain.

Radiative cooling includes H/He Lyman-$\alpha$ cooling, metal-line emission, and bremsstrahlung. Cooling rates are tabulated using \textsc{Chianti} \citep{Dere1997} for the dominant metal coolants C, N, O, Ne, Mg, Si, S, Ca, and Fe, assuming solar abundances \citep{Asplund2009}. To mimic large-scale heating processes that balance radiative losses in realistic halos, we implemented a mass-weighted heating prescription following \citet{McCourt2012}. In this approach, the net cooling losses are redistributed within horizontal layers, enforcing global thermal balance while allowing local thermal perturbations to grow and trigger TI within each layer. For the synthetic absorption analysis, we computed ion abundances in ionization equilibrium, including collisional ionization, radiative recombination, and photoionization. Ionization and recombination rate coefficients are obtained from the \textsc{Chianti} atomic database \citep{Dere2023} through the \textsc{ChiantiPy} Python interface \citep{Dere2013}. Photoionization cross sections are taken from \citet{Verner1996}, and the corresponding photoionization rates are computed for the adopted UV background radiation field \citep{Puchwein2019}. To account for attenuation in dense gas, we reduced the photoionization rates using the self-shielding prescription of \citet{Rahmati2013}.

\subsection{Cosmic-ray physics}
We modeled CRs using the two-moment CRMHD framework of \citet{Thomas2019}. The effective CR transport speed is regulated by interactions with Alfv\'en waves. When CRs stream faster than the local Alfv\'en speed, they excite resonant Alfv\'en waves via the gyroresonant instability \citep{Kulsrud1969, Shalaby2023, Lemmerz2025}. These waves scatter CRs, transferring momentum and energy and thereby limiting their transport speed. In the strong-scattering limit, CRs remain closely coupled to the waves and mainly stream along magnetic field lines at velocities near the local Alfv\'en speed. If scattering is weaker, CRs become less confined and their transport becomes more diffusive, which can be described by an effective diffusion coefficient. Cosmic rays can also exchange energy with the thermal gas. In addition to direct Coulomb and hadronic losses \citep{Enslin2007, Pfrommer2017}, CRs can heat the gas indirectly through the damping of self-generated Alfv\'en waves. In particular, non-linear Landau damping transfers wave energy to the plasma, reduces the wave amplitude, and lowers the CR scattering rate \citep{Miller1991}. In this work, we focused on this damping channel, which is expected to be the dominant one in CGM-like environments. For a more detailed discussion of the underlying physics, we refer the reader to \citet{Thomas2023}.

We also perform a set of simulations with purely diffusive CR transport, adopting a constant anisotropic CR diffusion coefficient, $\kappa_0$, along the mean magnetic field. In the two-moment formulation, such diffusion-only transport can be written in the form of the telegrapher's equation \citep{Gombosi1993, Litvinenko2013, Thomas2019},
\begin{equation}
\label{eq:telegraph}
    \tau_\mathrm{cr} \frac{\partial^2 \varepsilon_\mathrm{cr}}{\partial t^2} + \frac{\partial \varepsilon_\mathrm{cr}}{\partial t} = \bnabla\bcdot \left(\kappa_0 \bs{b}\bs{b}\bcdot\bnabla \varepsilon_\mathrm{cr} \right),
\end{equation}
where $\varepsilon_\mathrm{cr}$ is the CR energy density and $\bs{b}$ is the unit vector along the magnetic field. The parameter $\tau_\mathrm{cr}=3\kappa_0/c^2$ is the CR flux relaxation time. On timescales much longer than $\tau_\mathrm{cr}$, Eq.~\eqref{eq:telegraph} reduces to the usual anisotropic diffusion equation with diffusion coefficient $\kappa_0$. 

\begin{figure*}
    \centering
    \includegraphics[width=\linewidth]{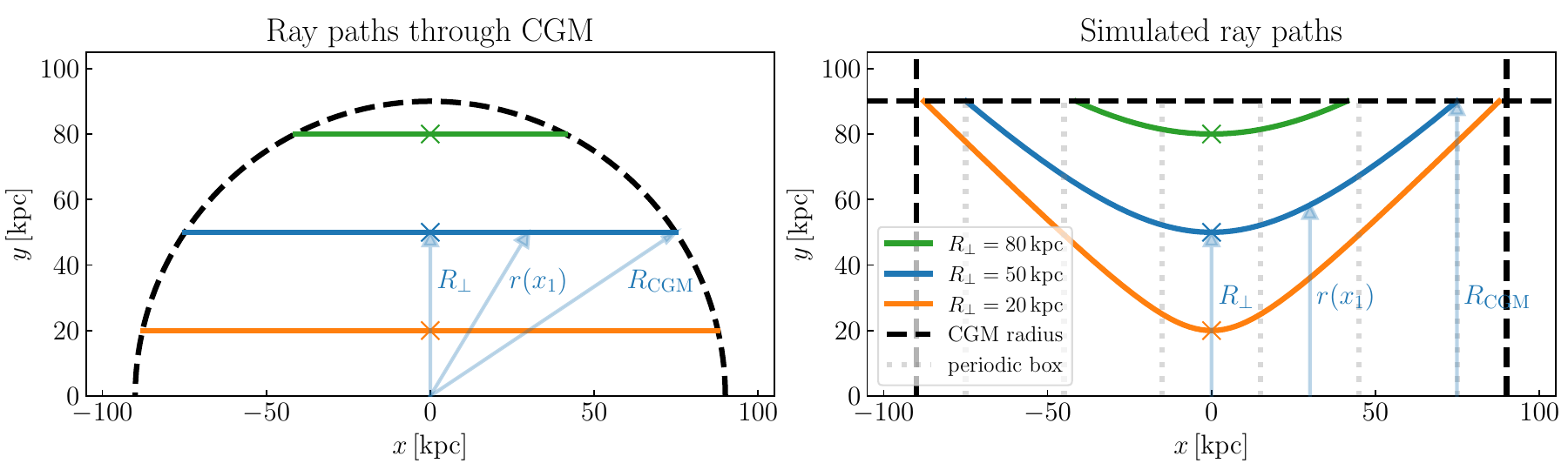}
    \caption{Geometry of the rays used to model absorption. 
    \textit{Left panel}: Example sightlines passing through a spherical halo of radius $R$. Each ray is characterized by its impact parameter $R_\perp$, defined as the perpendicular distance from the halo center to the ray's closest approach. The colored lines indicate different sightlines, and crosses mark their respective impact parameters. 
    \textit{Right panel}: Corresponding radial distance $R(x)$ 
    along the ray as a function of the line-of-sight coordinate $x$. Each ray is followed up to $x_\mathrm{max}=(R_\mathrm{CGM}^2-R_\perp^2)^{1/2}$ inside the halo, where the sightline intersects the simulated CGM boundary (dashed black lines). Vertical dotted lines indicate periodic replications of the simulation domain along the line of sight.
    }
    \label{fig:geometry}
\end{figure*}

\subsection{Initial conditions}
The initial gas distribution follows an isocooling atmosphere \citep{Butsky2020}, constructed such that the cooling time is constant with height. We set the ratio of cooling time to free-fall time to $t_\mathrm{cool}/t_\mathrm{ff} = 0.3$ at $|z| \sim 30\,\mathrm{kpc}$. This choice fixes the gravitational acceleration and scale height according to the smoothed vertical potential of \citet{McCourt2012} and is motivated by values found in cosmological simulations \citep[e.g.][]{Stern2021}. We neglected gas self-gravity, as it is not expected to be an important driver of cool-cloud condensation in the CGM conditions considered here \citep{Stern2016, Li2020}. To seed TI, we imposed an initial turbulent velocity field with a Kolmogorov power spectrum. The turbulence is normalized to a kinetic-to-thermal pressure ratio of $X_\mathrm{kin} = 0.3$. A divergence-free turbulent magnetic field is initialized with a mean magnetic-to-thermal pressure ratio of $X_\mathrm{mag} = 0.01$, corresponding to a weakly magnetized CGM.

We included CR pressure in the initial conditions by setting it to a fixed fraction of the thermal gas pressure in every computational cell, described by the parameter $X_\mathrm{cr} = P_\mathrm{cr}/P_\mathrm{th}$. Throughout this work, we focused on a CR-pressure-dominated halo with $X_\mathrm{cr}=3$. We explored three models for CR transport. In the first model, CRs are purely advected with the gas, corresponding to the limit of inefficient transport relative to the thermal plasma. In the second model, CRs undergo anisotropic diffusion along magnetic field lines with constant diffusion coefficients $\kappa_0 = 3 \times 10^{27}\,\mathrm{cm}^2\,\mathrm{s}^{-1}$ and $3 \times 10^{28}\,\mathrm{cm}^2\,\mathrm{s}^{-1}$. In the third model, we used the two-moment CR transport scheme, which evolves both the CR energy density and CR flux, allowing CRs to stream and diffuse relative to the gas. In these runs, CRs initially stream along magnetic field lines at the local Alfv\'en speed, while their effective transport speed and anisotropy evolve self-consistently.

We treat the CGM as isolated over timescales shorter than its characteristic dynamical, cooling, and accretion timescales. This approximation is most appropriate for studying local CGM processes over several hundred megayears, or over a few local cooling times. In this work, we analyze snapshots at $t=650\ \mathrm{Myr}$, corresponding to approximately six initial cooling times of the background atmosphere. Although this timescale is short compared to the characteristic CGM cycling times inferred from cosmological simulations, which show that CGM gas is continuously exchanged with the interstellar medium, satellites, and the intergalactic medium on timescales of order $(1$--$3)\,\mathrm{Gyr}$ \citep{Hafen2020}, it is well suited to an idealized post-outflow CGM state. In this interpretation, the halo has recently been enriched and energized by feedback-driven winds, as may occur following a burst of star formation. The subsequent evolution probes how local cooling, TI, and CR transport reshape this CGM before the longer timescale of cosmological cycling becomes important.

\subsection{Curved sightlines through a stratified CGM}
\label{sec:curved_rays}

To generate synthetic absorption spectra from our tall-box simulations, we constructed sightlines that approximate lines of sight through a galactic halo. The simulation domain is Cartesian and represents a vertically stratified patch of the CGM (cf. Fig.~\ref{fig:singleCloudZoom}), whereas observed absorption systems probe gas in an approximately spherical halo geometry. Therefore, directly tracing straight rays through the box would not appropriately capture the geometry of observed quasar sightlines, which are commonly characterized by their impact parameter, i.e., the projected distance between the background source and the galaxy center \citep{Tumlinson2017}. This geometric difference matters because both the density and velocity structure sampled along a sightline depend on the underlying halo geometry. In a spherical CGM, a sightline at fixed impact parameter passes through gas at continuously varying galactocentric radius. As a result, the local density, thermodynamic state, and projected line-of-sight velocity can change systematically along the ray, in particular if the velocity field is dominated by the radial direction, which is the case for TI. Straight rays through a Cartesian tall box, by contrast, probe the gas along a fixed direction and therefore miss part of the geometric projection effect that shapes observed absorption profiles.

To account for these effects, we constructed curved rays that approximate sightlines through a spherical halo at fixed impact parameter, $R_\perp$. For a sightline coordinate $x$, measured along the projected ray direction, the corresponding galactocentric radius is 
\begin{equation}
    r(x) = \sqrt{R_\perp^2 + x^2}.
\end{equation}
The part of the ray inside the halo is restricted to
\begin{equation}
    |x| \leq x_\mathrm{max} = \left(R_\mathrm{CGM}^2 - R_\perp^2\right)^{1/2},
\end{equation}
where $R_\mathrm{CGM}$ denotes the outer radius of the CGM.

Since the horizontal size of a single tall-box domain is smaller than the maximum chord length through the halo, we periodically replicated the box along the projected sightline direction. 
To ensure that even the longest sightlines, i.e. those with the smallest impact parameters, remain fully contained within the replicated domain, the number of copies is chosen such that
\begin{equation}
    N_\mathrm{copy} L \gtrsim 2 \sqrt{R_\mathrm{CGM}^2 - R_{\perp,\min}^2},
\end{equation}
where $L$ is the horizontal box size and $R_{\perp,\min}$ is the smallest impact parameter considered. With $N_\mathrm{copy}=7$, this condition is satisfied for our adopted radial range. The periodic copies should not be interpreted as independent halo volumes. Instead, they provide a controlled way to extend the stratified patch sufficiently far along the projected sightline so that the desired spherical geometry can be sampled without artificially truncating the ray. This construction ensures that each sightline samples the radial range appropriate for its impact parameter and reproduces the continuous change in projection angle expected for a spherical halo. Figure~\ref{fig:geometry} illustrates this geometry. The left panel shows representative sightlines intersecting the halo at different impact parameters, while the right panel shows the corresponding radial distance, $r(x)$, along the ray as mapped onto the replicated Cartesian domain. A detailed description of the ray construction through the Voronoi mesh is provided in Appendix~\ref{app:raytracing}.

We note, however, that this procedure remains an approximation. In particular, the use of periodic replicas means that long sightlines can encounter repeated realizations of the same stratified turbulent structure rather than independent halo regions. Moreover, the curved sightlines generally have longer path lengths through cool gas than straight rays through a spherical halo, which can increase the absolute column densities. These caveats should be kept in mind when interpreting absolute absorption strengths. Nevertheless, the same ray construction is applied to all CR transport models, so relative differences between models remain robust. We quantify the geometric contribution to the radial EW trends in Appendix~\ref{app:path_length_effect}. The curved-ray approach therefore provides a physically motivated compromise: it captures the dominant geometric and projection effects relevant for absorption-line formation in a stratified CGM, while retaining the controlled conditions and high resolution of the tall-box simulations.

\begin{figure}
    \centering
    \includegraphics[width=\linewidth]{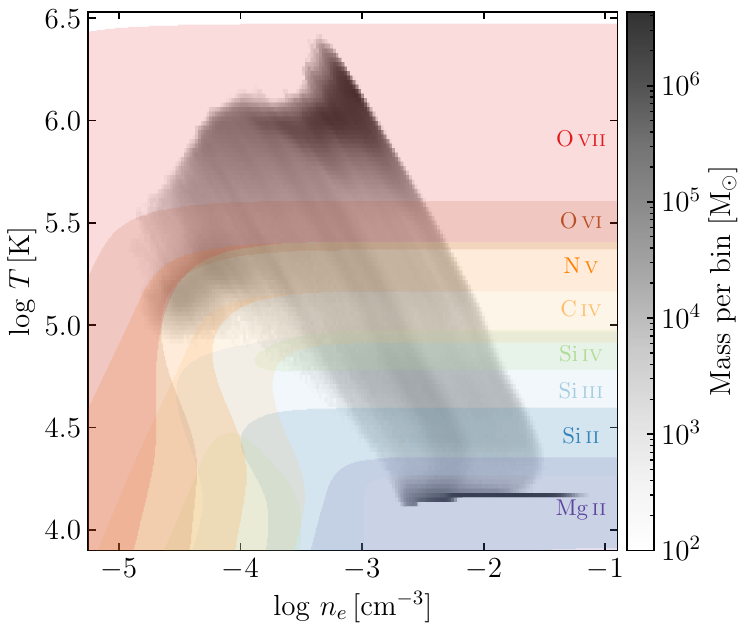}
    \caption{Ion fraction distribution for selected ions in our ionization model. Colored regions indicate where the fraction of each ion’s number density exceeds 0.1 of the corresponding element’s number density. The gray background shows the mass distribution of simulation cells.}
    \label{fig:tumlinson}
\end{figure}

\section{Observational diagnostics of the CGM}
\label{sec:theoreticalbackground}

In this section, we introduce the principal observational diagnostics employed throughout this study, which connect the physical properties of the simulated gas to measurable quantities. 
These diagnostics provide the basis for comparing the different CR transport models and for assessing how changes in the simulated CGM would appear in observations.


\subsection{Ion tracers of the CGM's multiphase structure}
The CGM is inherently multiphase, encompassing a wide range of densities and temperatures. It includes cool, dense clouds at $T \sim 10^4\,\mathrm{K}$, warmer transition regions around $T \sim 10^5\,\mathrm{K}$, and a hot, diffuse, volume-filling halo component with $T \gtrsim 10^6\,\mathrm{K}$ \citep{Tumlinson2017}. Because individual gas phases are difficult to isolate observationally, absorption lines of different ions serve as essential diagnostics of the thermodynamic structure of the CGM. Each ion has a characteristic ionization potential and peak ionization fraction, causing it to preferentially trace specific regions in the phase diagram \citep{Chen2026}. Figure~\ref{fig:tumlinson} shows the density–temperature diagram of one of our simulations with overlaid contours indicating where the ions \ion{Si}{II}, \ion{C}{IV}, \ion{O}{VI}, and \ion{O}{VII} reach significant abundance. The distribution reflects the combined effects of collisional ionization and photoionization in the halo environment.

Low-ionization species (e.g. \ion{Mg}{II}, \ion{Si}{II}), predominantly trace cool, relatively dense gas. These ions are abundant in cool clumps produced by TI and condensation processes \citep{McCourt2012, Sharma2012, Voit2017}. Their presence in the phase diagram is mostly confined to the higher-density, low-temperature region. Intermediate ions (e.g. \ion{Si}{III}, \ion{Si}{IV}, \ion{C}{IV}) probe warmer and more diffuse gas. \ion{Si}{III} typically occupies a regime overlapping with both the cold phase and the transition region toward warmer gas, making it sensitive to partially ionized envelopes surrounding cold clouds. \ion{Si}{IV} and \ion{C}{IV} trace gas at $T\sim10^{5}\,\mathrm{K}$, conditions commonly associated with turbulent mixing layers between cold clouds and the hot ambient medium. High-ionization species (\ion{O}{VI}, \ion{O}{VII}) trace progressively hotter and more diffuse gas. \ion{O}{VI} and \ion{O}{VII} both probe the warm-hot to hot, volume-filling halo gas, but they are sensitive to somewhat different temperature regimes. \ion{O}{VI} reaches its maximum abundance near $T\sim10^{5.5}\,\mathrm{K}$ and can therefore receive contributions from both the diffuse hot halo and interface layers surrounding cooler clouds \citep{Heckman2002, Werk2016, Oppenheimer2016}. \ion{O}{VII}, by contrast, peaks at $T\gtrsim10^6\,\mathrm{K}$ and more directly traces the hotter, volume-filling halo atmosphere.

While the phase diagram provides a thermodynamic classification of ion tracers, their spatial distribution within the halo further illustrates their diagnostic power. Figure~\ref{fig:singleCloudZoom} shows a thin projection ($d_\mathrm{proj}\sim 250$ pc) of the simulation box, with magnified views of a representative cold cloud on either side.  The panels on the left-hand side show hydrogen number density, $n_\mathrm{H}$, temperature, $T$, electron number density, $n_\mathrm{e}$, and turbulent velocity dispersion, $\sigma_{xy}$, demonstrating that the interior of a collapsed cloud is dense, cold, and less turbulent than the surrounding hot, dilute medium. The panels on the right display the number densities of different ions, namely $n_{\text{\ion{Mg}{II}}}$, $n_{\text{\ion{Si}{III}}}$, $n_{\text{\ion{C}{IV}}}$, and $n_{\text{\ion{O}{VII}}}$. These maps demonstrate that \ion{Mg}{II} is concentrated in the dense cloud core, confirming its role as a tracer of cold gas. \ion{Si}{III} extends beyond the core into the partially ionized envelope, while \ion{C}{IV} highlights the surrounding mixing layers where intermediate-temperature gas forms through shear, thermal exchange, and turbulent mixing. \ion{O}{VII} instead fills the diffuse background halo and avoids the cold structure, primarily tracing the hot ambient medium.

\subsection{Absorption-line spectroscopy as a diagnostic tool}
\label{subsec:absorption_line_spectroscopy}

Absorption-line spectroscopy is one of the most powerful methods for probing the gas in the CGM \citep{Savage1991}. Because absorption measurements are sensitive to much lower gas densities than emission measurements, the diffuse CGM is generally more readily detected in absorption than in emission \citep{Tumlinson2017}. Its physical properties are therefore commonly inferred from absorption features imprinted on the spectra of bright background sources, such as quasars or star-forming galaxies \citep{Rauch1998}. As the background radiation passes through the halo, ions in the intervening gas absorb photons at characteristic rest-frame wavelengths associated with bound--bound electronic transitions. Each absorption feature therefore selects gas containing a specific ion and encodes information about the amount, spatial distribution, and kinematics of that ion along the line of sight.

The basic quantity governing the attenuation of the background spectrum is the optical depth,
\begin{equation}
\label{eq:tau_integral}
    \tau_\lambda = \int n_\mathrm{ion}(s) \, \sigma_\lambda(s) \, \mathrm{d}s ,
\end{equation}
where $n_\mathrm{ion}$ is the number density of the absorbing ion, $\sigma_\lambda$ is the wavelength-dependent absorption cross section, and the integral is taken along the ray path, $s$. The transmitted flux is then
\begin{equation}
    F_\lambda = F_{\lambda,0} \exp(-\tau_\lambda),
\end{equation}
where $F_{\lambda,0}$ is the continuum flux. Absorption lines therefore appear where $\tau_\lambda$ becomes appreciable, with larger optical depths producing deeper absorption features.

The optical depth is closely related to the ion column density,
\begin{equation}
\label{eq:column_density}
    N_\mathrm{ion} = \int n_\mathrm{ion}(s) \, \mathrm{d}s ,
\end{equation}
which measures the total number of absorbing ions per unit area along the line of sight. However, the shape of an absorption line is not determined by column density alone. The absorption cross section depends on the oscillator strength, $f_\mathrm{osc}$, and the rest-frame wavelength, $\lambda_0$, of the transition \citep{Draine2011}, while thermal motions, natural broadening, turbulent motions, and line-of-sight bulk velocities distribute the absorption over wavelength or velocity space (see Appendix~\ref{app:absorptionlinesynthesis} for details). 

The Gaussian width of an individual absorption component is commonly characterized by the Doppler parameter $b$. If thermal and non-thermal broadening are both approximated as Gaussian, their contributions add in quadrature,
\begin{equation}
b^2 = b_\mathrm{th}^2 + b_\mathrm{nt}^2
= \frac{2 k_\mathrm{B} T}{m_\mathrm{ion}} + b_\mathrm{nt}^2 ,
\end{equation}
where $T$ is the gas temperature, $m_\mathrm{ion}$ is the ion mass, and $b_\mathrm{nt}$ represents unresolved turbulent or other non-thermal motions. A larger Doppler parameter broadens the central, approximately Gaussian part of the absorption profile and lowers its peak optical depth at fixed column density. 
Natural broadening, in contrast, produces a Lorentzian contribution to the profile. The combination of Doppler and natural broadening yields a Voigt profile, whose Lorentzian damping wings can become prominent far from the line center for transitions with sufficiently large optical depths and column densities \citep{Draine2011}.

For gas moving with a line-of-sight velocity $v_\mathrm{los}$, the absorption is Doppler-shifted relative to the rest wavelength. With the convention that positive $v_\mathrm{los}$ corresponds to blue-shifted absorption, this gives, to first order in
$v_\mathrm{los}/c$,
\begin{equation}
    \lambda_\mathrm{obs} \simeq \lambda_0 \left(1-\frac{v_\mathrm{los}}{c}\right),
\end{equation}
where $c$ is the speed of light. The observed absorption profile therefore contains information not only about the amount of absorbing material, but also about how this material is distributed in velocity space.

A useful observational measure of the total absorption strength is the EW,
\begin{equation}
    W_\lambda = \int \left(1 - F_\lambda/F_{\lambda,0}\right)\,\mathrm{d}\lambda,
\end{equation}
which quantifies the integrated flux decrement relative to the continuum. In the optically thin regime ($\tau_\lambda \ll 1$), the EW scales approximately as $W_\lambda \propto N_\mathrm{ion}$ \citep{Draine2011}. Thus, for weak lines, the EW provides a direct measure of the ionic column density. For saturated lines ($\tau_\lambda \gtrsim 1$), the above relation for $W_\lambda$ becomes non-linear in the column density, and the EW becomes increasingly sensitive to the velocity structure and unresolved sub-components of the absorbing gas, as described by the classical curve of growth \citep{Spitzer1978}. In this regime, increasing the velocity interval covered by optically thick absorption can increase the EW even if the ionic column density does not increase proportionally. At still larger column densities, the damping wings contribute substantially to the EW, causing it to increase again more strongly with column density.

\begin{figure*}
    \centering
    \includegraphics[width=\textwidth,height=0.8\textheight,keepaspectratio]{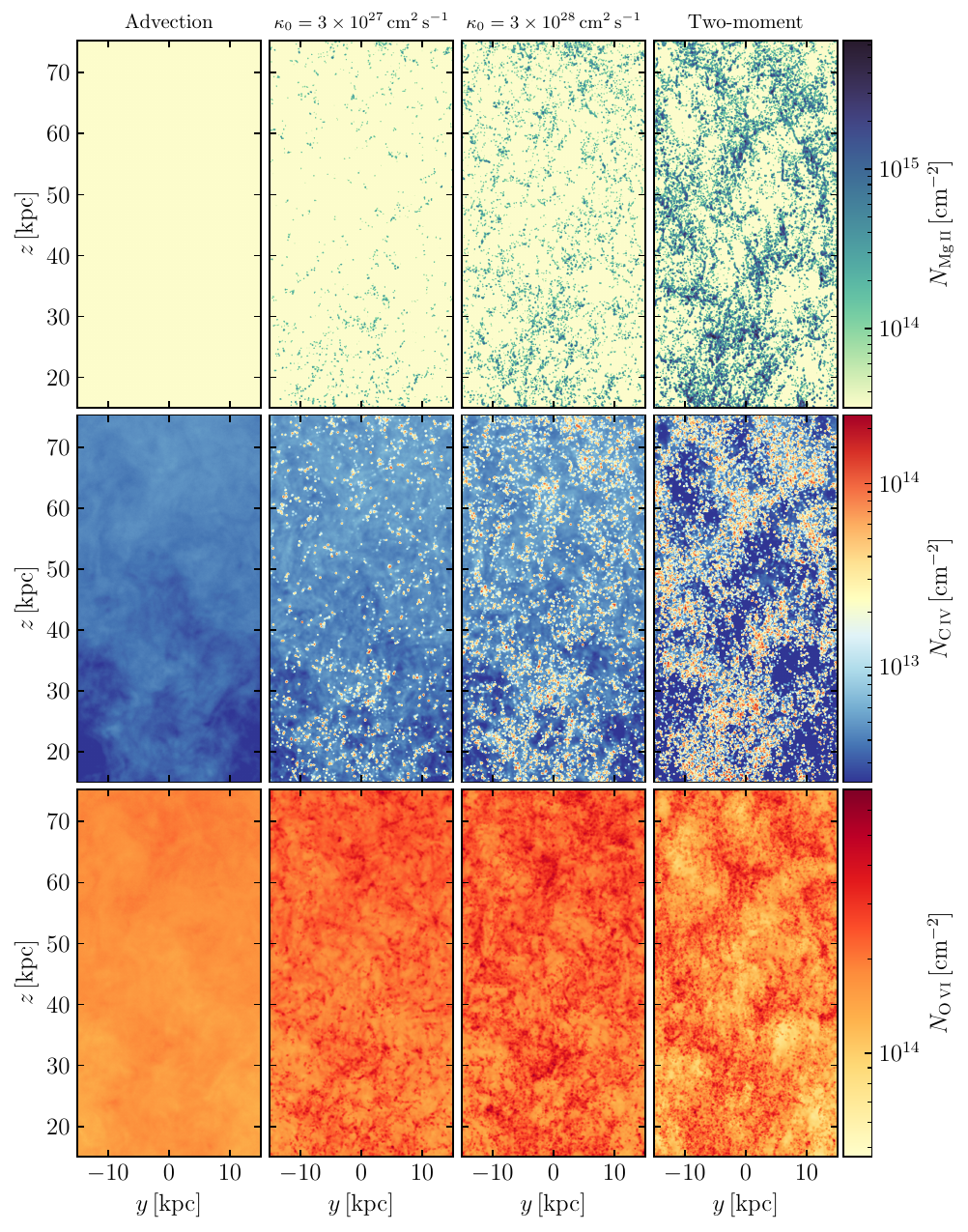}
    \caption{Column-density maps of \ion{Mg}{II}, \ion{C}{IV}, and \ion{O}{VI} (top to bottom), tracing cold, warm, and hot phases of the CGM, respectively, for simulations with different CR transport models. From left to right, the effective CR transport speed increases, ranging from pure advection (no active CR transport) to the two-moment transport model (fastest CR transport). Faster CR transport leads to systematically enhanced column densities of cold and warm gas, reflecting the increased prevalence of multiphase structure. In contrast, the spatial distribution of the hot phase remains broadly similar across models.
    }
    \label{fig:abundances}
\end{figure*}

\begin{figure*}
    \centering
    \includegraphics[width=\linewidth]{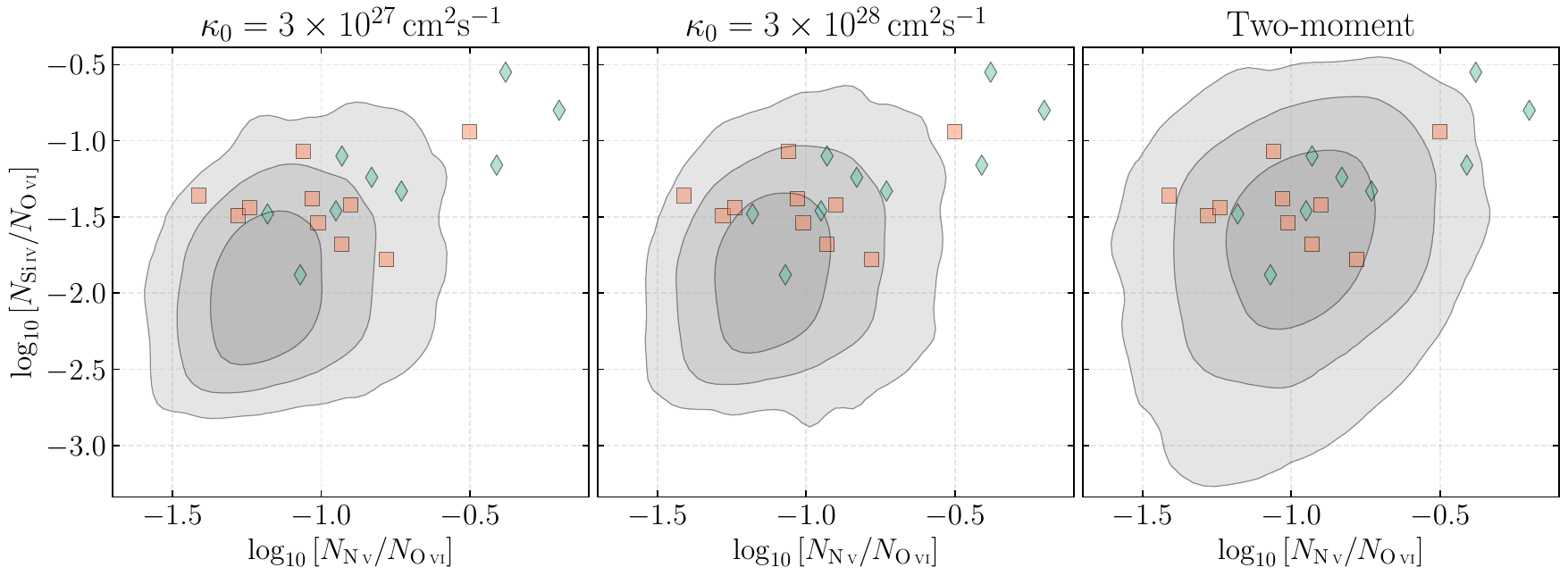}
    \caption{Abundance-ratio diagram for the different CR transport models. Each panel shows the relation between the logarithmic column-density ratios $\log(N_{\ion{N}{V}}/N_{\ion{O}{VI}})$ vs.\ $\log(N_{\ion{Si}{IV}}/N_{\ion{O}{VI}})$. The semi-transparent gray contours indicate the $1\sigma$, $2\sigma$, and $3\sigma$ confidence intervals of the simulated data which we restricted to typical detection limits for COS observations \citep[e.g.,][]{Werk2016}. Colored markers show observational measurements from \citet{Werk2016}, with orange squares denoting sightlines with broad \ion{O}{VI} absorption and green diamonds indicating systems with narrow \ion{O}{VI} components. The two-moment CR transport model appears to describe the observational data best.
    }
    \label{fig:abundanceratios}
\end{figure*}

Because low-ionization species predominantly trace cool, dense gas that is concentrated in compact clouds or cloud fragments, their absorption is therefore often confined to a small number of relatively narrow velocity components, producing sharp and localized features in the spectrum. Intermediate ions can arise both in warmer gas and in transition layers surrounding colder structures. They are therefore expected to show more complex profiles, combining absorption from compact interfaces with broader, lower-optical-depth contributions from spatially extended warm gas. High-ionization species trace hotter, more diffuse, and more volume-filling material. Since this gas typically extends over larger path lengths and samples a broader range of bulk motions, velocity gradients, and thermal velocities, high-ionization absorption profiles are expected to be broader and smoother than those of low-ionization species.

\ion{Mg}{II} is one of the most commonly used tracers of cool CGM gas \citep[e.g.,][]{Chen2010a, Werk2013, Huang2021}, but the strong \ion{Mg}{II} $\lambda\lambda2796,2803$ doublet is frequently saturated, both in observations \citep[e.g.][]{Tumlinson2017, Lan2018} and in our synthetic spectra. In this regime, the EW is no longer determined primarily by the column density, but also depends sensitively on the velocity spread and covering of absorbing gas along the line of sight. This point is particularly important for our idealized simulations. Since we do not include the full level of turbulence, galactic fountain motions, or feedback-driven dynamics expected in realistic inner CGM environments, the cool gas can produce narrow absorption components that saturate rapidly. 
The resulting \ion{Mg}{II} profiles should therefore not be interpreted as a direct prediction for observed line profiles but as qualitative tracers of cool-gas covering and kinematic spread, whereas observed saturated \ion{Mg}{II} systems can still contain substantial information through their resolved velocity structure, residual flux, and line-wing behavior.

To test whether our conclusions are driven by saturation effects, we therefore also analyze the much weaker \ion{Si}{II} $\lambda1808$ transition. Owing to its small oscillator strength, this line remains closer to the optically thin regime and provides a complementary diagnostic of cool, low-ionization gas. Although \ion{Si}{II} and \ion{Mg}{II} are not identical tracers, both probe similar gas phases, allowing us to assess whether the trends inferred from \ion{Mg}{II} persist for a substantially less saturated transition. A more detailed discussion of \ion{Mg}{II} saturation based on the doublet ratio is provided in Appendix~\ref{app:mgII_doublet_ratio}.

\section{Impact of CR transport on observational signatures}
\label{sec:results}

We now investigate how the transport of CRs modifies the observable properties of the CGM. We apply the diagnostics introduced above to quantify how different CR transport models affect the thermodynamic phase structure, ionic content, and absorption signatures of the halo gas.

Through their dynamical coupling to the plasma, CRs can substantially alter the thermal and kinematic state of the CGM. The strength and nature of this effect depend sensitively on how efficiently CR energy is transported through the halo \citep{Butsky2020, Tsung2023, Weber2025}. In models with inefficient transport, such as pure advection or slow diffusion, CRs remain more tightly coupled to the gas and can accumulate in overdense regions through adiabatic compression. The resulting enhanced CR pressure provides additional support against further compression, thereby suppressing TI and reducing the formation of cold clouds. In this regime, the CGM remains comparatively smooth, with weaker density contrasts and a lower cold-gas fraction \citep{Butsky2020, Weber2025}.

By contrast, efficient CR transport redistributes CR energy more effectively throughout the halo and prevents excessive CR pressure build-up in dense regions. This allows overdense gas to cool and condense more readily, leading to stronger density contrasts between cold clumps and the surrounding hot medium \citep{Weber2025}. In addition, CR pressure gradients can drive gentle gas motions, buoyancy, and mixing, which may enhance the formation of intermediate-temperature interface gas. The resulting halo develops a richer multiphase structure, consisting of cold cloud cores, warm transition layers, and an extended hot ambient medium.

These transport-dependent changes should leave distinct imprints on observable CGM diagnostics. If CR pressure suppresses TI, low-ionization tracers associated with cold gas are expected to weaken, while the velocity structure of the absorbing material becomes less complex. Conversely, when efficient CR transport promotes the formation of multiphase structure, the column densities, CFs, EWs, and velocity widths of low- and intermediate-ionization species can increase. High-ionization tracers, which mainly probe the diffuse volume-filling halo gas, are expected to respond more weakly. 

\subsection{Column densities of tracer ions}
\label{columndensitiesofiontracers}

We first examine how CR transport affects the spatial distribution of ions that trace different thermodynamic phases of the CGM. Since different ions are sensitive to different temperature regimes, their column densities provide a useful first diagnostic of how CR transport redistributes gas between phases. Figure~\ref{fig:abundances} shows column-density maps of representative cool-, warm-, and hot-gas tracers, given by \ion{Mg}{II}, \ion{C}{IV}, and \ion{O}{VI}, respectively.

For the cool phase, with characteristic temperatures of $T\sim10^4\,\mathrm{K}$, models with faster CR transport exhibit systematically higher \ion{Mg}{II} column densities. This trend is consistent with more efficient TI when CRs can escape from collapsing overdensities. Faster CR transport reduces the build-up of CR pressure support inside dense gas, allowing overdense regions to cool and condense more effectively \citep{Weber2025}. The resulting increase in cool gas mass directly enhances the column densities of low-ionization species.

We find a similar trend for the intermediate-temperature gas traced by \ion{C}{IV}, which typically arises at around $T\sim10^5\,\mathrm{K}$. The enhanced \ion{C}{IV} column densities in the faster-transport models are likely linked to the increased formation of cool structures. This interpretation is supported by the spatial correlation between regions of enhanced \ion{C}{IV} column density and the \ion{Mg}{II}-bearing cool gas. As more cool gas condenses out of the hot halo, the total surface area of cool--hot interfaces increases, producing more transition-layer gas in which \ion{C}{IV} can reach high ion fractions \citep{Slavin1993, Ji2019}. The increase in \ion{C}{IV} therefore appears to be an indirect consequence of enhanced cool-gas formation and the associated growth of mixing layers around condensed clouds.

In contrast, the high-ionization tracer \ion{O}{VI}, associated with gas at $T\gtrsim10^{5.5}\,\mathrm{K}$, shows smaller systematic differences between the CR transport models. The hot halo remains volume filling and is comparatively insensitive to the range of CR transport efficiencies explored here. Although condensation removes some material from the hot phase, the global thermodynamic structure of the hot gas is largely preserved. Consequently, the column densities of high-ionization tracers vary only weakly between the different transport models.

The pure-advection model forms little to no cool gas and therefore does not produce appreciable low-ionization absorption. We therefore exclude this model from the following absorption-line analysis and focus on the transport models that generate a detectable multiphase CGM.

\begin{figure*}
    \centering
    \includegraphics[width=\linewidth]{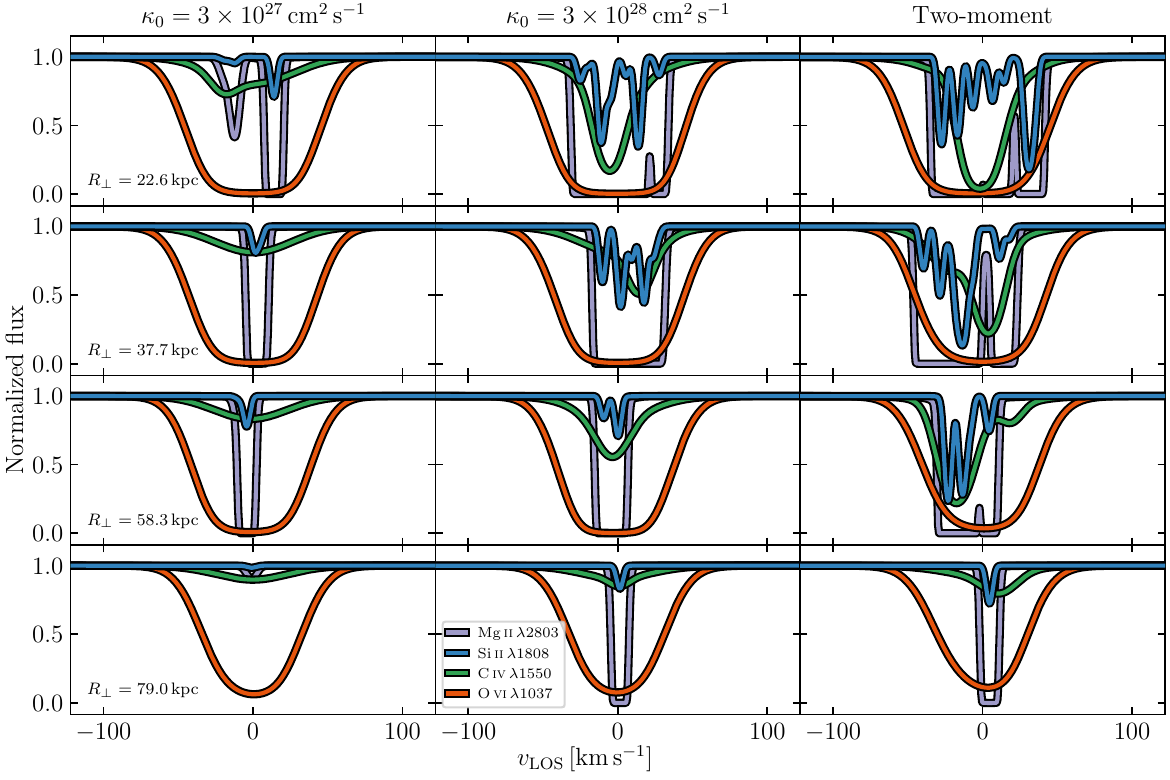}
    \caption{Synthetic absorption spectra along representative sightlines. Rows correspond to sightlines at different impact parameters, while columns show the different CR transport models. For a given row, the impact parameter, $R_\perp$, and transverse position, $y$, are kept fixed across all simulations, such that each row compares identical sightline geometries between the different transport models. In each panel, we plot the residual flux profiles of \ion{Mg}{II}, \ion{Si}{II}, \ion{C}{IV}, and \ion{O}{VI}. Variations in the CR transport speed primarily affect the absorption strength, component structure, and velocity extent of the cool and warm gas, whereas the hot phase traced by \ion{O}{VI} remains comparatively insensitive to the transport model.
    }
    \label{fig:single_line_gallery}
\end{figure*}

\begin{table}
    \centering
    \begin{tabular}{c|c|c|c|c}
        Ion & $\lambda_0$ [\AA] & $f_\mathrm{osc}$ &  $\gamma_\lambda\,[\mathrm{s^{-1}}]$ & $T_\mathrm{peak}$ [K]\\ 
        \hline
        $\ion{Mg}{II}$ & $2796.352$ & 0.608   & $2.60\times10^{8}$  &$\sim 10^{4}~~$  \\
        $\ion{Mg}{II}$ & $2803.531$ & 0.303   & $2.57\times10^{8}$  &$\sim 10^{4}~~$  \\
        $\ion{Si}{II}$ & $1808.013$ & 0.00249 & $2.54\times10^{6}$  & $\sim 10^{4}~~$ \\
        $\ion{C}{IV}$  & $1550.784$ & 0.0952  & $2.64\times10^{8}$  &$\sim 10^{5}~~$  \\
        $\ion{O}{VI}$  & $1037.615$ & 0.066   & $4.09\times10^{8}$  &$\sim 10^{5.5}$ \\

    \end{tabular}
    \caption{Atomic data for the absorption lines used in this work. Listed are the rest-frame wavelength $\lambda_0$, the oscillator strength $f_\mathrm{osc}$, the radiative damping constant $\gamma_\lambda$, and the characteristic temperature $T_\mathrm{peak}$ at which each ion reaches its maximum ion fraction under CIE.     
    }
    \label{tab:ions}
\end{table}

\begin{figure*}
    \centering
    \includegraphics[width=\linewidth]{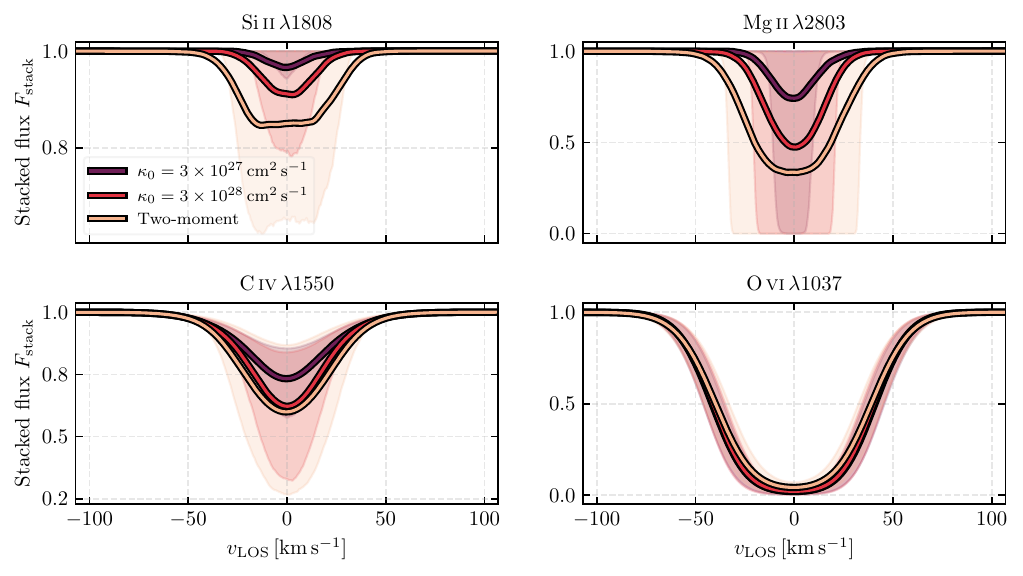}
    \caption{Stacked absorption spectra of \ion{Si}{II} $\lambda 1808$, \ion{Mg}{II} $\lambda 2803$, \ion{C}{IV} $\lambda 1550$, and \ion{O}{VI} $\lambda 1037$ for different CR transport models. Solid lines show the normalized flux calculated from the mean residual flux across all sightlines, while shaded regions indicate the 16th–84th percentile range. Variations in CR transport speed strongly affect both the depth and width of low- and intermediate-ionization absorption lines, whereas higher-ionization tracers exhibit comparatively weaker changes. Note the different $y$-axis ranges.
    }
    \label{fig:spectra_3panel_transport}
\end{figure*}

\subsection{Abundance ratios}
\label{sec:abundanceratios}

While absolute column densities provide valuable information about the total amount of absorbing material along a line of sight, ratios of ion column densities offer a more robust diagnostic of the physical state of the CGM. Absolute column densities depend not only on the thermodynamic state of the gas, but also on metallicity, line-of-sight path length, and the total gas mass. These quantities are often uncertain observationally. In contrast, column-density ratios partly remove these normalization dependencies and are therefore less sensitive to the total metal content or geometric assumptions. 

Because different ions occupy different regions of density--temperature space, their relative abundances probe the distribution of gas across multiple phases \citep[e.g.][]{Wakker2012, Tumlinson2017}. We focus on ratios involving \ion{O}{VI}, \ion{N}{V}, and \ion{Si}{IV}, which together trace the balance between cool, transition-temperature, and hot gas. The \ion{N}{V}/\ion{O}{VI} ratio is sensitive to the temperature and ionization state of warm-hot gas, since both ions are abundant in highly ionized plasma but peak under somewhat different conditions. The \ion{Si}{IV}/\ion{O}{VI} ratio instead compares cooler, lower-ionization material to the hotter or more extended gas traced by \ion{O}{VI}. Higher \ion{Si}{IV}/\ion{O}{VI} values therefore indicate a larger contribution from cool or interface gas, while lower values suggest that the absorption is dominated by hotter, more diffuse halo gas \citep{Shull2011, Wakker2012}.

In Fig.~\ref{fig:abundanceratios}, we show the distribution of these column-density ratios for the different CR transport models. Each panel shows the relation between $\log(N_{\ion{N}{V}}/N_{\ion{O}{VI}})$ and $\log(N_{\ion{Si}{IV}}/N_{\ion{O}{VI}})$, illustrating how CR transport modifies the relative contributions of cool, warm, and hot gas phases in the CGM. The semi-transparent gray contours denote the $1\sigma$, $2\sigma$, and $3\sigma$ confidence intervals of the simulated distributions, restricted to typical observational limits for these ions \citep[e.g.][]{Werk2016}. Observational measurements from \citet{Werk2016} are overplotted as colored markers, with orange squares denoting sightlines with broad \ion{O}{VI} absorption and green diamonds denoting systems with narrow \ion{O}{VI} components.

The different CR transport models show different levels of agreement with the observed abundance-ratio distribution. While the slow-transport models are broadly consistent with the observations at the $3\sigma$ level, the two-moment CR transport model provides a closer match: most of the observational data points lie within the $1\sigma$ confidence interval of the simulated distribution. This indicates that the efficient CR transport model better reproduces the typical observed combination of $N_{\ion{N}{V}}/N_{\ion{O}{VI}}$ and $N_{\ion{Si}{IV}}/N_{\ion{O}{VI}}$. The abundance-ratio comparison therefore supports the interpretation that CR transport efficiency affects not only the total amount of cool gas, but also the relative contribution of cool, transition-temperature, and hot gas phases to the observable CGM.

\subsection{Absorption line analysis}
\label{sec:absorption_line_analysis}

Having characterized the ionic column densities, we now examine the corresponding synthetic absorption spectra. We first discuss individual sightlines to illustrate how different CR transport models can affect absorption line strength, component structure, and velocity extent, before using stacked spectra to assess the statistical robustness of these trends. The atomic data for the transitions considered in this work are listed in Table~\ref{tab:ions}. All of them are ground-state resonance lines commonly used in absorption-line studies of circumgalactic and intergalactic gas \citep{Cashman2017}. Excited-state transitions are not included, as their level populations are expected to be negligible in the low-density CGM gas considered here \citep{Draine2011}.

Figure~\ref{fig:single_line_gallery} presents synthetic absorption spectra for individual sightlines at four representative impact parameters and for the different CR transport models. The columns correspond to the different transport models, while the rows show sightlines at increasing impact parameter. In each panel, we show the residual flux profiles of \ion{Mg}{II}, \ion{Si}{II}, \ion{C}{IV}, and \ion{O}{VI} in purple, blue, green, and red, respectively.

\begin{figure*}
    \centering
    \includegraphics[width=0.49\linewidth]{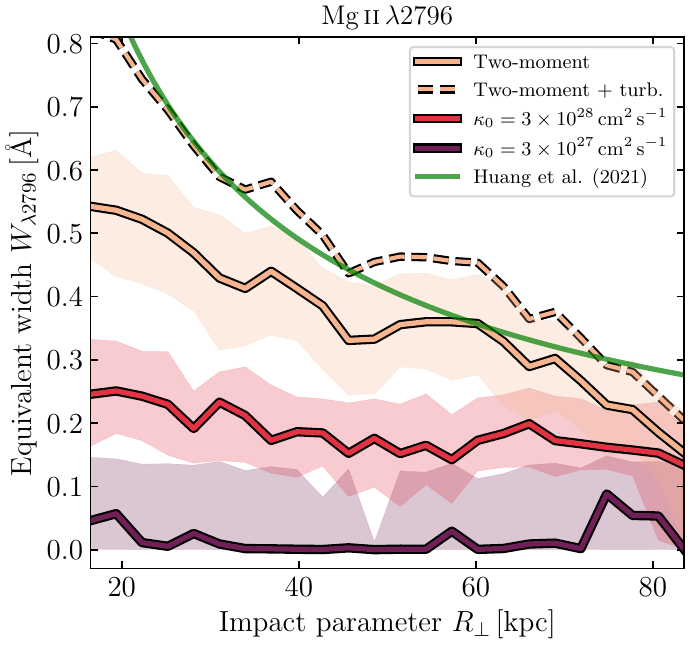}
    \includegraphics[width=0.49\linewidth]{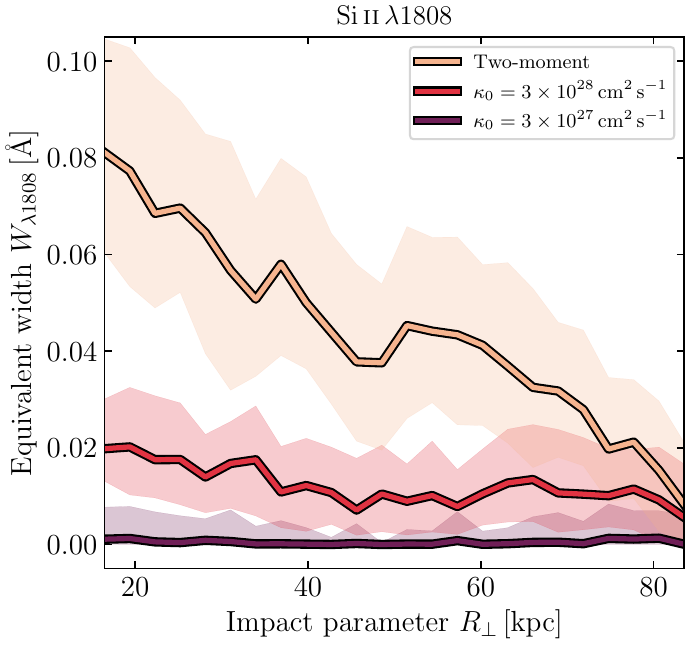}
    \caption{Equivalent width of \ion{Mg}{II} (left) and \ion{Si}{II} (right) as a function of impact parameter for the different CR transport models. Solid lines denote the median EW in each radial bin, while the semi-transparent shaded regions show the corresponding 30th--70th percentile range. In the left panel, the green dashed line shows the observational relation from \citet{Huang2021}. The dashed line illustrates the result of the turbulence model, in which an unmodeled Doppler broadening term with $b_\mathrm{turb}(R_\perp)=20\,\mathrm{km\,s^{-1}}(z/15\,\mathrm{kpc})^{-1}$ is applied to the \ion{Mg}{II} spectra of the two-moment run.
    }
    \label{fig:ew_vs_impactparameter}
\end{figure*}

The spectra reveal several qualitative trends. Absorption from the cool-gas tracers \ion{Mg}{II} and \ion{Si}{II} becomes systematically stronger with increasing CR transport speed, largely independent of impact parameter. For \ion{Mg}{II}, which is often saturated, this increase mainly appears as broader absorption rather than substantially deeper line cores. The weaker \ion{Si}{II} transition, by contrast, reveals a more detailed velocity structure: in the faster-transport models, individual sightlines intersect a larger number of distinct absorption components. These components are typically narrow, with individual velocity widths of $\lesssim 10\,\mathrm{km\,s^{-1}}$. The \ion{Mg}{II} profiles are generally aligned in velocity with the \ion{Si}{II} absorption, confirming that both ions trace the same cool, low-ionization structures. However, the \ion{Mg}{II} lines are saturated in many cases, so their residual flux no longer directly reflects variations in column density. We interpret the increasing strength and complexity of the low-ion absorption as a consequence of enhanced cold-gas formation in models with more efficient CR transport, which increases the probability that a given sightline intersects multiple kinematically distinct structures. 

Within a given CR transport model, the \ion{Si}{II} and \ion{C}{IV} absorption tends to weaken with increasing impact parameter. This trend follows from the declining gas density with radius and the shorter path length through the halo at larger $R_\perp$. Both effects reduce the ion column density along the sightline and therefore lead to weaker absorption features at larger impact parameters.

The intermediate-temperature ion tracer \ion{C}{IV} exhibits trends that are broadly similar to those of \ion{Si}{II}, but with important differences. As for the cooler gas, the \ion{C}{IV} absorption strengthens with increasing CR transport speed. In the faster-transport models, the \ion{C}{IV} features are often aligned in velocity with the \ion{Si}{II} components, suggesting that a significant fraction of the \ion{C}{IV} arises in transition layers or interfaces surrounding the cool clouds (cf. Fig.~\ref{fig:singleCloudZoom}). The individual \ion{C}{IV} components are, however, broader than the corresponding \ion{Si}{II} features. This is expected because \ion{C}{IV} traces warmer gas and because the lower atomic mass of carbon compared to silicon leads to larger thermal Doppler broadening. In the slower-transport models, by contrast, the \ion{C}{IV} profiles appear smoother and are less clearly associated with individual \ion{Si}{II} components. This suggests that in these models, the \ion{C}{IV} absorption is then dominated less by compact transition layers around cold structures and more by spatially extended warm gas. The changing morphology of the \ion{C}{IV} profiles therefore indicates a shift in the dominant origin of the absorption.

The hotter gas traced by \ion{O}{VI} exhibits only minor variations between the different CR transport models. This reflects that \ion{O}{VI} predominantly probes the diffuse, volume-filling halo gas, whose thermodynamic and kinematic structure is comparatively weakly affected by the changes in CR transport explored here. We find only a mild decrease in \ion{O}{VI} absorption strength with increasing impact parameter. As for the other ions, this radial trend can be attributed to the declining gas density and the shorter path length through the halo at larger $R_\perp$.

To move from individual sightlines to the statistical properties of the full sample, we next consider stacked absorption profiles. Individual spectra highlight the diversity of absorption components and velocity structures, whereas stacked spectra provide a more
robust measure of the average absorption strength and kinematic width produced by each CR transport model.

Figure~\ref{fig:spectra_3panel_transport} shows the stacked synthetic absorption spectra for the different CR transport models. Solid lines denote the average spectra defined below, and the semi-transparent regions indicate the corresponding 16th--84th percentile
range. The top row shows the stacked residual-flux profiles of the low-ionization tracers \ion{Si}{II} and \ion{Mg}{II}. The bottom row presents \ion{C}{IV} and \ion{O}{VI}, which trace intermediate- and high-ionization gas, respectively.
The stacked profiles are computed from the average residual flux across all sightlines. Specifically, we define
\begin{equation}
    F_\mathrm{stack} = \langle F_0\exp\left(- \tau \right)\rangle,
\end{equation}
where $\langle ... \rangle$ denotes the mean over the full sightline sample. 
Each spectrum is constructed from 5600 sightlines sampling impact parameters between $15$ and $85\,\mathrm{kpc}$. For each impact parameter, the transverse position, $y$, is varied between $-10\,\mathrm{kpc}$ and $+10\,\mathrm{kpc}$. Both dimensions are sampled with a spatial step size of approximately $0.5\,\mathrm{kpc}$. 

The stacked spectra confirm the trends inferred from individual sightlines. For \ion{Mg}{II} and \ion{Si}{II}, faster CR transport produces absorption profiles that are both deeper and broader compared to pure-diffusion models. The increased depth reflects the larger low-ionization column densities already identified in Fig.~\ref{fig:abundances}, while the enhanced width indicates that the additional cool gas is distributed over a wider range of line-of-sight velocities. This broadening is likely caused by a combination of turbulent and bulk motions. More efficient CR transport promotes the formation of a larger number of cool clouds and interfaces, increasing the amount of mixing between cool condensed gas and the surrounding hot halo. These interfaces can drive shear motions and small-scale velocity dispersion, thereby increasing the turbulent contribution to the line width. At the same time, the condensed structures are denser than their surroundings and can decouple partially from the background flow, making them more susceptible to gravitational infall. As a result, individual sightlines intersect cool structures with a broader range of bulk line-of-sight velocities.

The response of \ion{C}{IV} follows the same qualitative trend, but with a smaller amplitude. In the faster-transport models, the stacked \ion{C}{IV} profiles become stronger and moderately broader, consistent with the enhanced association between \ion{C}{IV} and the cold-cloud population seen in individual spectra. Conversely, the weaker and smoother profiles in the slower-transport models support the interpretation that \ion{C}{IV} is then less dominated by compact cloud interfaces and instead receives a larger relative contribution from spatially extended warm gas. The stacked \ion{O}{VI} profiles vary only weakly across the CR transport models. Both their depth and width remain broadly similar, indicating that the \ion{O}{VI}-bearing gas is not strongly reshaped by the changes in CR transport.

\subsection{Equivalent widths}
\label{sec:equivalent_widths}
The EW of an absorption line provides an integrated diagnostic of the amount and kinematic distribution of absorbing gas along a given line of sight \citep[e.g.][]{Chen2010a, Bordoloi2011, Nielsen2013a, Richter2016}. It measures the total absorption strength integrated over wavelength. In the optically thin regime, the EW scales directly with the ionic column density, $N_\mathrm{ion}$. For stronger or partially saturated lines, however, the EW also becomes sensitive to the velocity structure of the absorbing material, because absorption distributed over a broader velocity interval spans a wider wavelength range. The EW therefore encodes both the amount of absorbing gas and its kinematic complexity.

Figure~\ref{fig:ew_vs_impactparameter} shows the EW of \ion{Mg}{II} and \ion{Si}{II} as a function of impact parameter for the different CR transport models. We divided the radial range from $15\,\mathrm{kpc}$ to $85\,\mathrm{kpc}$ into 20 equally spaced bins and plot the resulting statistics at the corresponding bin centers. In each radial bin, we computed the median EW using only sightlines with $W_\lambda > 0$. The shaded regions indicate the width of the EW distribution within each bin, bounded by the 30th and 70th percentiles. For \ion{Mg}{II}, we additionally compared our synthetic EWs to the observational relation from \citet{Huang2021}, specifically their best-fit model for isolated star-forming galaxies. 

The different CR transport models produce clear systematic differences in the \ion{Mg}{II} EWs. The two-moment transport model yields the strongest absorption across the full radial range, with median values reaching $W_\lambda \simeq 0.6\,\text{\AA}$ at small impact parameters and declining gradually to $\sim 0.15\,\text{\AA}$ at larger radii. In contrast, the constant-diffusion models produce substantially weaker absorption. For $\kappa_0 = 3\times 10^{28}\,\mathrm{cm^2\,s^{-1}}$, the typical EWs lie in the range $0.1\,\text{\AA} \lesssim W_\lambda \lesssim 0.3\,\text{\AA}$, whereas the $\kappa_0 = 3\times 10^{27}\,\mathrm{cm^2\,s^{-1}}$ model remains below $W_\lambda \lesssim 0.15\,\text{\AA}$ and has median values close to zero over much of the radial range. Both constant-diffusion models show only a weak dependence on impact parameter.

To test whether these trends are driven by saturation of the strong \ion{Mg}{II} transition, the right panel of Fig.~\ref{fig:ew_vs_impactparameter} shows the corresponding EW--impact-parameter relation for the much weaker \ion{Si}{II} $\lambda1808$ line. The \ion{Si}{II} EWs are significantly smaller, as expected from its lower oscillator strength, but they exhibit the same qualitative dependence on CR transport. This indicates that the enhanced absorption in the two-moment model is not solely an artifact of \ion{Mg}{II} saturation, but reflects a genuine increase in the amount and/or kinematic complexity of cool, low-ionization gas.

The simulated \ion{Mg}{II} EWs at small impact parameters fall below the observed relation from \citet{Huang2021}. This is likely a consequence of the idealized nature of our setup. Although our simulations are designed to represent CR-pressure-dominated halos with $X_\mathrm{cr}=3$, as expected in star-forming galaxies where supernova-driven feedback injects CRs that can be transported into the CGM by winds \citep[e.g.][]{Ji2020, Hopkins2021a, Butsky2022, Thomas2023, Thomas2025b}, we do not explicitly model the central galaxy or the associated stirring of the inner halo. Processes such as continuous feedback-driven galactic winds and fountain flows, as well as cosmological accretion can drive turbulence and broaden the velocity distribution of cool gas in the inner CGM \citep[e.g.][]{Lan2014, Chen-ShiFan2017}. Because saturated \ion{Mg}{II} EWs are highly sensitive to this velocity structure, the lack of driven turbulence in our simulations likely suppresses the absolute EW values at small impact parameters.

We recomputed the \ion{Mg}{II} EWs with a simplified turbulence model and added a non-thermal Doppler broadening term to the optical-depth profiles. 
Motivated by the expectation that turbulent stirring driven by stellar feedback is strongest in the inner CGM, we adopted a radially declining velocity broadening of the form
\begin{equation}
    b_\mathrm{turb}(R_\perp) = b_0 \left( z/z_0 \right)^{-1} ,
    \label{eq:b_turb}
\end{equation}
with $b_0=20\,\mathrm{km\,s^{-1}}$ and $z_0=15\,\mathrm{kpc}$. This gives the largest additional broadening at small impact parameters and reduces its effect in the outer halo. 

This model leaves the gas distribution unchanged and only mimics unmodelled small-scale velocity dispersion in the \ion{Mg}{II} optical-depth profiles. Since additional broadening approximately conserves the EW of optically thin lines, its effect is largest for saturated \ion{Mg}{II}, where the EW depends on the velocity extent of optically thick absorption. This model therefore estimates how much of the central EW deficit could be attributed to missing inner-CGM turbulence. We show the result of this experiment for the two-moment model in the left panel of Fig.~\ref{fig:ew_vs_impactparameter} as dashed lines. We find that this modest additional broadening increases the central \ion{Mg}{II} EWs and brings the efficient-transport model closer to the observed relation, supporting the interpretation that the remaining discrepancy at small impact parameters is at least partly caused by the lack of turbulent stirring in the inner CGM in our idealized simulations.

\subsection{Covering fractions}

\begin{figure}
    \centering
    \includegraphics[width=\linewidth]{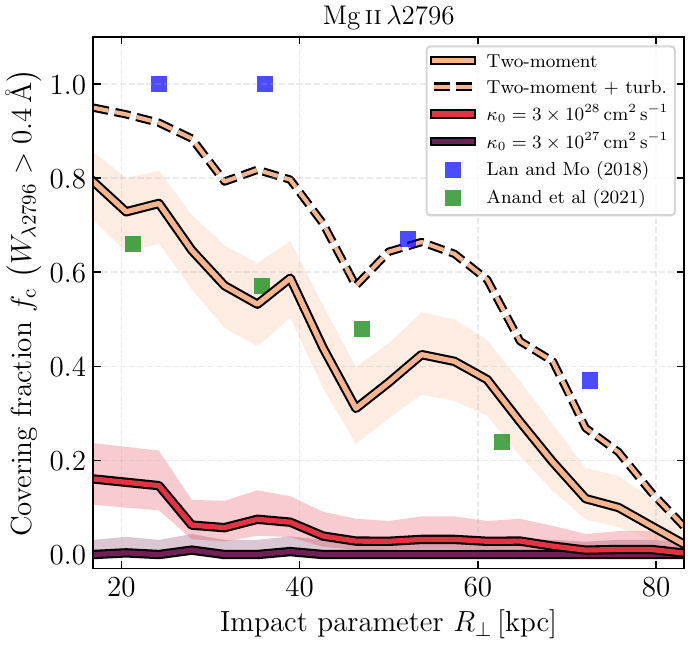}
    \caption{ 
    Covering fraction of \ion{Mg}{II} as a function of impact parameter for the different CR transport models. Solid lines denote the CF in each radial bin, while the semi-transparent shaded regions indicate the corresponding 99\% Wilson confidence intervals. We additionally show observational measurements from \citet{Lan2018} and \citet{Anand2021} as blue and green squares, respectively. The dashed line illustrates the result of the turbulence model for the two-moment CR transport simulation, using the same prescription as in Fig.~\ref{fig:ew_vs_impactparameter}.
    }
    \label{fig:coveringfraction}
\end{figure}

The CF of low-ionization absorbers provides a direct measure of the spatial prevalence of cool gas in galaxy halos and is therefore a sensitive diagnostic of the multiphase structure of the CGM \citep{Bordoloi2014, Rubin2015}. We computed the \ion{Mg}{II} $\lambda2796$ CF as a function of impact parameter for our suite of simulations with different CR transport models. Each sightline through the simulation box is treated as a potential detection, and we classified a sightline as detectable if its rest-frame EW exceeds $W_{\lambda2796} > 0.4\,\text{\AA}$. This threshold is chosen to match the observational measurements used for comparison below.

In each impact-parameter bin, the CF is defined as $f_\mathrm{c}=N_\mathrm{det}/N_\mathrm{tot}$, where $N_\mathrm{det}$ is the number of sightlines satisfying the EW criterion and $N_\mathrm{tot}$ is the total number of sightlines in that bin. We divided the radial range from $15\,\mathrm{kpc}$ to $85\,\mathrm{kpc}$ into 20 equally spaced bins and plot the resulting CFs at the corresponding bin centers. The CF is thus defined as the fraction of sightlines that would show substantial absorption in a fiducial observation. For our simulations and assuming a uniform distribution of absorbers along individual line of sights, the CF is a proxy of how much of the probed simulation volume contains a significant amount of \ion{Mg}{II}. To quantify the statistical uncertainty associated with the finite number of sightlines, we computed Wilson binomial confidence intervals for each radial bin and show them as shaded regions.

Figure~\ref{fig:coveringfraction} presents the resulting \ion{Mg}{II} CFs for the different CR transport models. The two-moment CR transport model produces the largest spatial extent of detectable \ion{Mg}{II} absorption, with CFs reaching $f_\mathrm{c}\sim0.8$ at small impact parameters and decreasing gradually toward zero at the largest radii. In contrast, the constant-diffusion models yield substantially lower CFs. For $\kappa_0=3\times10^{28}\,\mathrm{cm^2\,s^{-1}}$, $f_\mathrm{c}$ decreases from ${\sim}0.2$ in the inner halo to nearly zero at large impact parameters. The slowest transport model, with $\kappa_0=3\times10^{27}\,\mathrm{cm^2\,s^{-1}}$, produces only sparse detectable absorption, with $f_\mathrm{c}\simeq0$ across most of the radial range and only a weak radial dependence.

We compare our simulated \ion{Mg}{II} CFs to observational measurements for star-forming galaxies from \citet{Lan2018} and \citet{Anand2021}, using the same EW threshold of $W_{\lambda2796}>0.4\,\text{\AA}$. The efficient CR transport model broadly reproduces the elevated \ion{Mg}{II} CFs reported by \citet{Anand2021} over much of the radial range, although it remains below the measurements of \citet{Lan2018} by a factor of ${\sim}1.5$--$2$. This offset may partly reflect differences in halo sample selection and analysis methods \citep{Anand2021}, since CFs are known to depend on galaxy population and star-formation activity \citep{Lan2018}. As discussed above, missing inner-CGM turbulence can also reduce the EWs of saturated \ion{Mg}{II} absorption and therefore lower the CF measured above a fixed EW threshold. We show the CF obtained after applying the turbulence model (Eq.~\ref{eq:b_turb}) to the two-moment run as a dashed line in Fig.~\ref{fig:coveringfraction}. This additional broadening increases the CF, especially at small and intermediate impact parameters, and moves the model into the range bracketed by the observational measurements of \citet{Anand2021} and \citet{Lan2018}. In contrast, the slower CR transport models produce substantially lower CFs, indicating that they do not generate a sufficiently extended distribution of cool, low-ionization gas above the applied threshold. The $\kappa_0=3 \times 10^{28} \, \mathrm{cm^2\,s^{-1}}$ model underpredicts the observed CFs by approximately an order of magnitude, while the slowest transport model, with $\kappa_0=3\times10^{27}\,\mathrm{cm^2\,s^{-1}}$, falls short by more than an order of magnitude over most of the radial range. 

\subsection{Velocity width diagnostics}
\label{velocity_width_diagnostics}

To quantify the kinematic extent of the absorption features in more detail, we employed the velocity width statistic $v_{90}$, defined as the velocity interval enclosing 90\% of the total optical depth \citep[e.g.][]{Prochaska1997, Ellison2006, Zou2021, Carr2025}. Unlike single-component line-width measures such as the Doppler parameter $b$ or the full width at half maximum, $v_{90}$ is sensitive to the full structure of the absorption profile because it captures contributions from thermal broadening, turbulent motions, bulk flows, and in particular from the presence of multiple absorbing components along the line of sight. We computed the cumulative optical depth along the line of sight,
\begin{equation}
    C(v) = \frac{\int^{v}_{-\infty} \tau(v')\,\mathrm{d}v'}{\int^{+\infty}_{-\infty} \tau(v')\,\mathrm{d}v'},
\end{equation}
and define
\begin{equation}
    v_{90} = v_{95} - v_{5},
    \label{eq:v90}
\end{equation}
where $v_{5}$ and $v_{95}$ correspond to the velocities at which $C(v_{5})=0.05$ and $C(v_{95})=0.95$, respectively. By construction, $v_{90}$ captures the velocity range over which the bulk of the absorbing material is distributed.

\begin{figure}
    \centering
    \includegraphics[width=\linewidth]{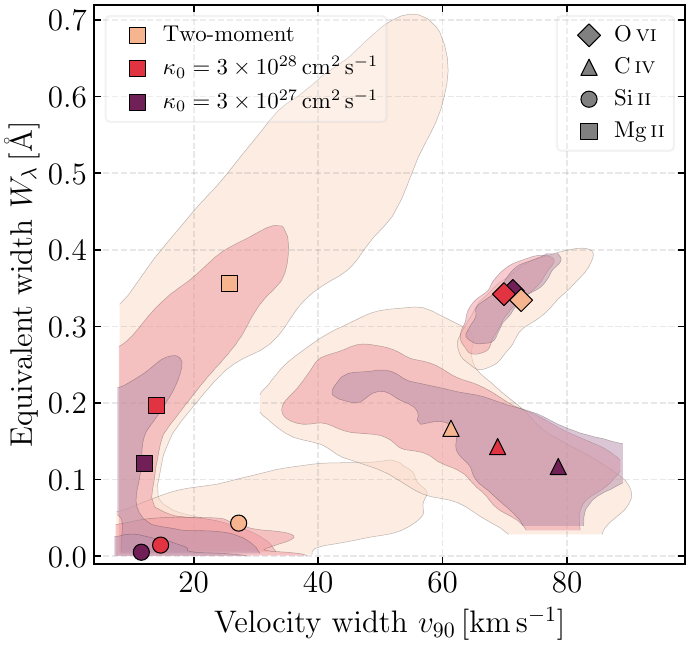}
    \caption{Equivalent width as a function of velocity width $v_{90}$ (Eq.~\ref{eq:v90}) for different ions and CR transport models. Large symbols indicate median values across all sightlines, while the filled contours enclose approximately 85\% of the smoothed distribution for each ion and CR transport model. The distribution highlights how absorption strength correlates with kinematic complexity for different CR transport models.
    }
    \label{fig:v90_vs_ew}
\end{figure}

Figure~\ref{fig:v90_vs_ew} shows the relation between the velocity width, $v_{90}$, and the EW, $W_\lambda$, for different ions and CR transport models. Large symbols indicate the median values computed from the full sightline sample, while the filled contours enclose approximately 85\% of the smoothed distribution. Colors distinguish the different CR transport models, while symbol shapes denote the ion species: squares for \ion{Mg}{II}, circles for \ion{Si}{II}, triangles for \ion{C}{IV}, and diamonds for \ion{O}{VI}. In the following analysis, we only included sightlines with $W_\lambda>10^{-5}\,\text{\AA}$. Since the individual points shown in the figure represent only a random subset, the sightlines contributing to the median values are not necessarily all visible in the plotted sample.

The different ion species occupy distinct regions of the $W_\lambda$--$v_{90}$ plane. In general, warmer gas is expected to exhibit larger velocity widths partly because the thermal contribution to the line profile increases with temperature. Differences in $v_{90}$ between ions therefore reflect both the kinematic structure of the absorbing gas and the temperature-dependent thermal broadening of the corresponding transitions.
\ion{Si}{II} is mainly confined to the low-$W_\lambda$, low-$v_{90}$ regime, with $W_\lambda < 0.2\,\text{\AA}$ and $v_{90}\simeq (7$--$90)\,\mathrm{km\,s^{-1}}$. \ion{Mg}{II} reaches substantially larger EWs, up to $W_\lambda \simeq 1\,\text{\AA}$, while spanning a similar range in velocity width. This supports the interpretation that \ion{Mg}{II} and \ion{Si}{II} trace the same cool, low-ionization gas, with the larger \ion{Mg}{II} EWs primarily reflecting its stronger transition and frequent saturation. \ion{C}{IV} occupies an intermediate region of the diagram, with $0.02\,\text{\AA} \lesssim W_\lambda < 0.5\,\text{\AA}$ and $v_{90}\simeq (30$--$110)\,\mathrm{km\,s^{-1}}$, consistent with its role as a tracer of warmer gas and transition layers around cool structures. By contrast, \ion{O}{VI} predominantly lies in the high-$W_\lambda$, high-$v_{90}$ regime, with $0.2\,\text{\AA} \lesssim W_\lambda < 0.4\,\text{\AA}$ and $v_{90}\simeq (60$--$105)\,\mathrm{km\,s^{-1}}$, reflecting its association with more extended, hotter gas that samples a broader range of line-of-sight velocities.

We now focus on how the locations of different ions in the $W$--$v_{90}$ plane change with CR transport. These trends reveal that CR transport does not affect all ionization phases in the same way. 
For the low-ionization tracers \ion{Mg}{II} and \ion{Si}{II}, both $v_{90}$ and $W$ increase with increasing CR transport speed, producing an approximately linear, positive relation between EW and $v_{90}$. This indicates that more efficient CR transport gives rise to stronger and kinematically more extended low-ionization absorption. 
For the intermediate-ionization tracer \ion{C}{IV}, the trend is qualitatively different. As the CR transport speed decreases, the sightlines develop an anti-correlation between $W$ and $v_{90}$: the absorption extends over a broader velocity interval, while the total EW decreases. This suggests that the dominant \ion{C}{IV} contribution is redistributed from compact, high-column transition layers around cold structures to a more extended, lower-density warm phase (cf. Fig.~\ref{fig:abundances}). While interface-dominated \ion{C}{IV} produces relatively strong absorption over a limited velocity range, the extended warm gas samples a broader range of line-of-sight velocities but contributes a lower optical depth per velocity interval. Broader \ion{C}{IV} profiles therefore do not necessarily imply stronger absorption, but instead trace a transition from small-scale transition-layer absorption toward weaker, more diffuse, and kinematically extended warm gas (cf. Fig.~\ref{fig:single_line_gallery}). 
In contrast, the high-ionization tracer \ion{O}{VI} shows only a weak sensitivity to the CR transport model, with both $v_{90}$ and $W$ remaining broadly similar across the different runs. This indicates that the warm-hot gas traced by \ion{O}{VI} is less directly affected by the variations in CR transport.

\section{Discussion and conclusions}
\label{sec:discussion}

\subsection{Limitations}

While our simulations provide valuable insight into how CR transport can affect observables in a multiphase CGM, several limitations should be kept in mind when interpreting the results.

Our setup models an idealized vertical CGM column rather than a fully cosmological halo. This allows for high spatial resolution and a controlled comparison of CR transport models, but it omits important aspects of realistic galaxy environments, including cosmological inflows, mergers, halo-to-halo variations, and the time-dependent coupling between disk feedback and halo gas recycling. In particular, the simulations do not include explicit star formation, supernova- or CR-driven winds, or other small-scale interstellar medium feedback processes that would continuously inject mass, metals, turbulence, and CRs into the halo. The resulting gas distribution and kinematics should therefore be interpreted as a simplified framework designed to isolate the effects of CR transport, rather than as a fully representative model of the CGM in real galaxies.

Our ionization modeling assumes a uniform solar metallicity throughout the halo. This simplification neglects spatial metallicity variations, which can affect the absolute ion column densities and EWs. We compute ion abundances in ionization equilibrium, including collisional ionization, radiative recombination, and photoionization by the UV background, with attenuation in dense gas accounted for using a self-shielding prescription. Under this assumption, efficiently cooling gas approaches an equilibrium state near $T\sim10^4\,\mathrm{K}$, where several low-ionization species can become abundant under the conditions considered here. However, this treatment does not include time-dependent ionization or full radiative transfer. Non-equilibrium effects may be important for rapidly cooling gas, for example if recombination lags behind the local cooling time, thereby modifying the ion fractions relative to the equilibrium expectation. Consequently, the absolute absorption properties of the coolest gas should be interpreted with caution. Since the same metallicity and ionization assumptions are applied to all simulations, however, the relative differences between the CR transport models remain informative.

A related limitation is the finite resolution of our simulations. Intermediate-temperature ions (\ion{C}{IV}, \ion{Si}{IV}, \ion{N}{V}) often trace transition-temperature gas in mixing layers around cool clouds. In these interface regions, the detailed ion fractions are sensitive to the resolved temperature and density structure. In our simulations, the cooling length of this mixing-layer gas is typically sampled by roughly one computational cell (cf. \citet{Weber2025}), implying that these interfaces are marginally captured rather than fully resolved. We therefore interpret the intermediate-ionization columns primarily as tracers of relative differences between CR transport models. A dedicated resolution study would be required to quantify the convergence of these interface-sensitive ion columns.

\subsection{Conclusions}

In this paper, we investigated how different CR transport models shape the observable absorption-line signatures of CR-pressure-dominated CGMs. Using synthetic spectra derived from high-resolution simulations, we examined how CR transport modifies the multiphase structure of the halo gas and how these changes translate into typical CGM observables. Our main findings are as follows:

\begin{enumerate}

    \item The effective CR transport speed has a strong impact on the multiphase structure of the CGM. More efficient CR transport leads to a substantially larger amount of cool ($T\sim10^4\,\mathrm{K}$) and warm ($T\sim10^5\,\mathrm{K}$) gas, consistent with enhanced thermal instability and condensation when CRs can escape from collapsing overdensities (cf. Fig.~\ref{fig:abundances}). This change in phase structure is also reflected in the abundance-ratio diagram: the two-moment transport model produces $N_{\ion{N}{V}}/N_{\ion{O}{VI}}$ and $N_{\ion{Si}{IV}}/N_{\ion{O}{VI}}$ ratios that more closely overlap with the observed distribution, indicating a more realistic balance between cooler, transition-temperature, and hot gas phases (cf. Fig.~\ref{fig:abundanceratios}).

    \item These changes in phase structure are directly reflected in the synthetic absorption spectra. Faster CR transport produces systematically deeper and broader absorption in low- and intermediate-ionization tracers (cf. Fig.~\ref{fig:spectra_3panel_transport}). The increased absorption strength reflects the larger columns of cool and warm gas, while the broader profiles indicate that the absorbing material spans a wider range of line-of-sight velocities. This broadening arises from the larger number of kinematically distinct cool structures intersected by each sightline, together with the enhanced turbulent and bulk motions of the multiphase gas (cf. Fig.~\ref{fig:v90_vs_ew}).

    \item The radial EW profiles provide a clear observational diagnostic of CR transport. The two-moment transport model produces the strongest \ion{Mg}{II} and \ion{Si}{II} absorption over the full radial range, whereas the constant-diffusion models yield substantially weaker EWs (cf. Fig.~\ref{fig:ew_vs_impactparameter}). The low-ion EW profiles therefore demonstrate that the imprint of CR transport is not restricted to individual sightlines, but persists in the global radial distribution of absorption strengths. 

    \item The \ion{Mg}{II} CFs increase strongly with CR transport efficiency (cf. Fig.~\ref{fig:coveringfraction}). For an EW threshold of
    $W_{\lambda2796}>0.4\,\text{\AA}$, the two-moment model reaches high CFs in the inner halo and lies within the range of observationally inferred \ion{Mg}{II} CFs for star-forming galaxies. In contrast, the slower CR transport models underpredict the observed CFs by approximately an order of magnitude or more, indicating that they do not generate a sufficiently extended distribution of cool, low-ionization gas.

    \item The physical origin of the \ion{C}{IV}-bearing warm gas changes with CR transport efficiency. In the slow-transport models, \ion{C}{IV} absorption is mainly associated with extended, large-scale warm gas in the halo. In the efficient-transport model, by contrast, enhanced cool-cloud formation increases the surface area of cool--hot interfaces, so that a larger fraction of the \ion{C}{IV} absorption arises in mixing layers around condensed structures (cf. Fig.~\ref{fig:abundances}). This shift is also visible in the $W_\lambda$--$v_{90}$ plane: slow transport tends to produce low-EW, high-$v_{90}$ \ion{C}{IV} absorbers, whereas efficient transport gives rise to stronger \ion{C}{IV} absorption with smaller velocity widths (cf. Fig.~\ref{fig:v90_vs_ew}). Thus, CR transport can affect not only the amount of intermediate-temperature gas, but also its spatial origin and connection to the cool phase.

    \item Higher-ionization gas responds more weakly to changes in CR transport. In particular, the \ion{O}{VI} absorption profiles show only modest variations in EW and velocity width across the transport models (cf. Fig.~\ref{fig:v90_vs_ew}). This indicates that CR transport primarily regulates the cool, condensed phase and the associated transition layers, while the more diffuse, volume-filling hot halo is comparatively less affected. 

\end{enumerate}

Overall, our results demonstrate that CR transport physics can leave clear and measurable imprints on the absorption properties of CR-pressure-dominated galaxy halos. Efficient CR transport promotes the formation of cool and warm gas, increases the strength and CF of low-ionization absorption, and changes the physical origin of intermediate-ionization material. These diagnostics therefore provide a powerful set of complementary observables for constraining the still poorly understood transport properties of CRs in galactic halos.

\begin{acknowledgements} 
MW, TT, and CP acknowledge support by the European Research Council under ERC-AdG grant PICOGAL-101019746. TU acknowledges support by the European Research Council under ERC-AdG grant SPECMAP-CGM-101020943. This work was supported by the North-German Supercomputing Alliance (HLRN) under project bbp00070. The projections and slices presented in this work were generated using \texttt{Paicos} \citep{Berlok2024}. Additionally, all Python-based analysis scripts relied on \texttt{matplotlib} \citep{Matplotlib}, \texttt{seaborn} \citep{Seaborn} for visualization including color maps, and \texttt{numpy} \citep{Numpy} for numerical computations. Atomic line data were obtained from the NIST Atomic Spectra Database \citep{NIST_ASD}.
\end{acknowledgements}

\bibliographystyle{aa_url}
\bibliography{refs}

\begin{appendix}


\section{Ray-tracing algorithm}
\label{app:raytracing}

We compute sightlines through an unstructured three-dimensional Voronoi mesh defined by a set of mesh-generating points $\{\bs{x}_i\}$. In contrast to grid-based ray-tracing methods that rely on explicitly stored cell connectivity, our approach is mesh-less in the sense that no precomputed topological information about the Voronoi tessellation is required. Instead, all geometric relations are reconstructed on-the-fly from the positions of the mesh-generating points.

The Voronoi tessellation is uniquely defined by the set of mesh-generating points, with each cell consisting of all spatial locations closer to its generating point than to any other. We exploit this defining property to determine cell transitions dynamically. In practice, all spatial queries are performed using a KD-tree \citep{SciPy}, which enables efficient nearest-neighbor searches for both cell identification and neighbor selection.

The algorithm supports both straight rays (\ref{app:straightray}) and curved rays (\ref{app:curvedray}). In both cases, the ray is propagated iteratively from cell to cell until it either leaves the computational domain or a maximum number of steps is reached. At each step, candidate interfaces are constructed from a finite set of nearby mesh-generating points, and the next intersection is determined geometrically.

\subsection{Straight ray propagation}
\label{app:straightray}

We define a ray by its starting position $\bs{x}_0$ and a normalized direction $\hat{\bs{n}}_{\rm ray}$. The trajectory is given by
\begin{equation}
\label{eq:raystart}
\bs{x}(s) = \bs{x}_0 + s\,\hat{\bs{n}}_{\rm ray},
\end{equation}
where $s$ denotes the path length along the ray.

The initial Voronoi cell is determined via a KD-tree nearest-neighbor query, identifying the mesh-generating point closest to $\bs{x}_0$. For the current cell with center $\bs{x}_i$, we select a finite set of $k$ nearest mesh-generating points $\{\bs{x}_j\}$ using the KD-tree. This set approximates the true Voronoi neighbors. To ensure forward propagation, we restrict to candidates satisfying
\begin{equation}
(\bs{x}_j - \bs{x}_i) \bs\cdot \hat{\bs{n}}_{\rm ray} > 0.
\end{equation}

The interface between two Voronoi cells is approximated by the perpendicular bisector of their mesh-generating points. For each candidate neighbor $j$, we define the midpoint
\begin{equation}
\bs{x}_{\rm mid} = \bs{x}_i + \frac{1}{2}(\bs{x}_j - \bs{x}_i),
\end{equation}
and the corresponding unit normal vector
\begin{equation}
\hat{\bs{n}}_{ij} = \frac{\bs{x}_j - \bs{x}_i}{|\bs{x}_j - \bs{x}_i|}.
\end{equation}
The interface plane is then given by
\begin{equation}
\hat{\bs{n}}_{ij} \bs\cdot (\bs{x} - \bs{x}_{\rm mid}) = 0.
\end{equation}

Substituting Eq.~\eqref{eq:raystart} yields the path length to the intersection with this interface,
\begin{equation}
s_{ij} =
\frac{\hat{\bs{n}}_{ij} \bs\cdot (\bs{x}_{\rm mid} - \bs{x}_0)}
{\hat{\bs{n}}_{ij} \bs\cdot \hat{\bs{n}}_{\rm ray}}.
\end{equation}
We retain only candidates with $s_{ij} > 0$ and non-vanishing denominators. The smallest valid value determines the next interface crossing.

In addition to Voronoi interfaces, we also compute the distance to the boundary of the computational domain. The next event along the ray is given by the smaller of the two distances, i.e. either a cell-to-cell transition or an exit from the domain. This ensures that the final partial path through the last intersected cell is properly accounted for.

The ray position is updated to the intersection point, and the procedure is repeated iteratively until no valid intersection is found, the ray exits the domain, or a predefined maximum number of steps is reached. We note that restricting the interface search to a finite number of nearby mesh-generating points introduces a controlled approximation, which we verify to be negligible for the adopted neighbor counts.

\begin{figure}[!t]
    \centering
    \includegraphics[width=\linewidth]{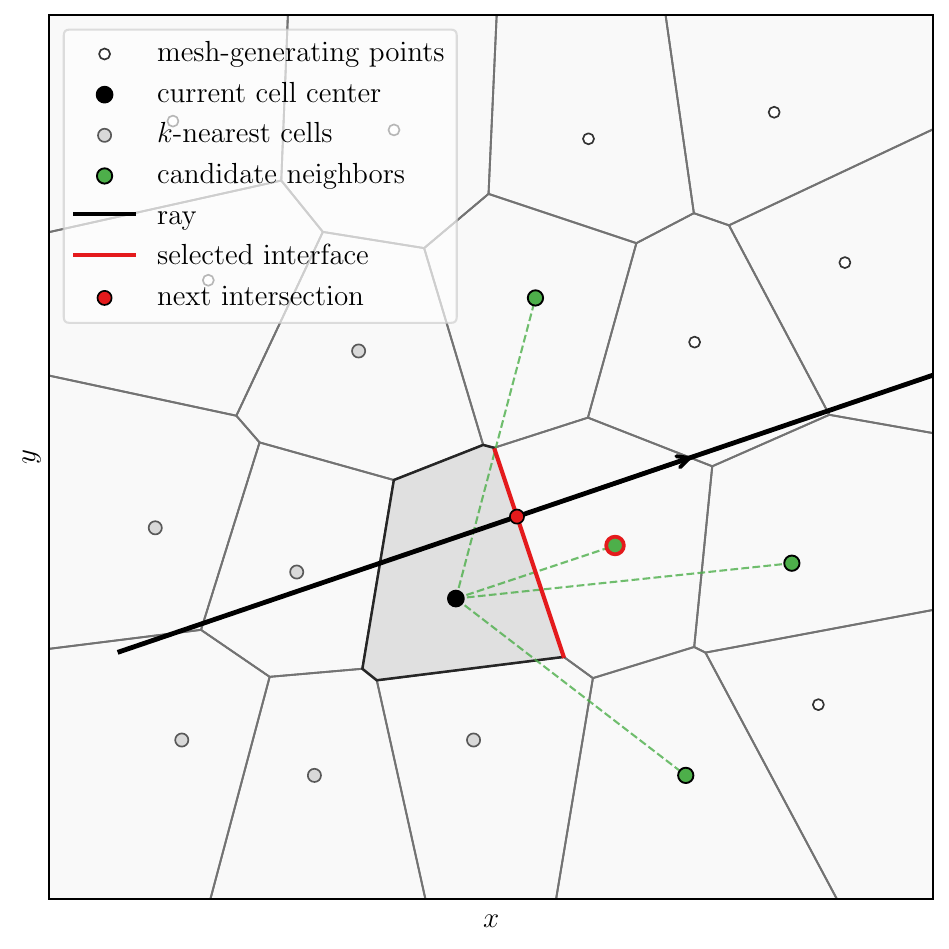}
    \caption{Schematic illustration of the ray-tracing algorithm on a Voronoi mesh. The mesh-generating points $\bs{x}_i$ define the Voronoi tessellation, with one cell (shaded) highlighted as the current cell. A set of nearby mesh-generating points is identified via a KD-tree search (grey points), from which candidate neighbors are selected by requiring a positive projection along the ray direction (green points). The corresponding candidate interfaces are given by the perpendicular bisectors between the current cell and its neighbors (one example is shown in red). The next cell transition is determined by computing the intersection of the ray with these interfaces and selecting the closest valid intersection.}
    \label{fig:raytracing}
\end{figure}

\subsection{Curved ray propagation}
\label{app:curvedray}

To model curved ray trajectories, we describe the ray by a parametric curve $\bs{x}(t)$ with an initial position and direction consistent with Eq.~\eqref{eq:raystart}. Within each Voronoi cell, the curve is advanced using a sub-stepping procedure. Starting from the current position, the curve is incremented in small parametric steps $\Delta t$. In our implementation, $\Delta t$ is chosen as a fixed fraction of a characteristic minimum cell size, ensuring that cell transitions are resolved even in highly irregular regions of the mesh.

At each sub-step, a new point $\bs{x}(t+\Delta t)$ is evaluated, and a KD-tree query is used to determine the index of the Voronoi cell containing this point. As long as the cell index remains unchanged, the curve is still inside the current cell and the integration continues. When a change in the cell index is detected, a boundary crossing has occurred. The local propagation direction of the ray is then approximated by the tangent of the curve, computed from successive sub-step positions,
\begin{equation}
\hat{\bs{n}}_{\rm tan} = \frac{\Delta \bs{x}}{|\Delta \bs{x}|}.
\end{equation}
Using this local tangent, the precise intersection with the Voronoi interface is reconstructed via the same plane–ray intersection formalism as in the straight-ray case. This procedure is repeated iteratively, yielding a piecewise reconstruction of the curved trajectory through the Voronoi mesh. While the method approximates the true curve locally as linear within each cell, it accurately captures both the geometric path and the sequence of traversed cells.

\section{Absorption line synthesis}
\label{app:absorptionlinesynthesis}

We compute synthetic absorption spectra by integrating the frequency-dependent optical depth along each ray through the simulation domain \citep[cf.][]{Draine2011}. For each cell $i$ intersected by a sightline, we evaluate the Voigt absorption profile based on the local thermodynamic and kinematic properties of the gas.

The Doppler-shifted line center in each cell is given by
\begin{equation}
\nu_i = \nu_0 \left(1 - \varv_{\mathrm{LOS},\,i}/c\right),
\label{eq:nu_i}
\end{equation}
where $\nu_0$ is the rest-frame frequency of the transition and $\varv_{\mathrm{LOS},\,i}$ is the line-of-sight velocity of the gas. The absorption profile of each cell is modeled as a normalized Voigt profile,
\begin{equation}
    \phi_{\nu,i}
    =
    \frac{H(a_i,u_i)}
         {\sqrt{\pi}\,\Delta\nu_{\mathrm{D},i}},
\end{equation}
where the dimensionless Voigt function is a convolution of a Gaussian (describing thermal and turbulent broadening) and a Lorentzian profile (describing the broadening due to a finite life time of an excited state) and is given by 
\begin{equation}
    H(a_i,u_i)
    =
    \frac{a_i}{\pi}
    \int_{-\infty}^{\infty}
    \frac{\exp(-x^2)}
         {(u_i-x)^2+a_i^2}
    \,\mathrm{d}x .
\end{equation}
The associated dimensionless variables are given by
\begin{equation}
u_i = \frac{\nu - \nu_i}{\Delta \nu_{\mathrm{D},\,i}}, 
\qquad
a_i = \frac{\Gamma}{4\pi \Delta \nu_{\mathrm{D},\,i}},
\label{eq:voigt_params}
\end{equation}
where $\Gamma$ is the natural damping constant of the transition and $\Delta \nu_{D,i}$ is the Doppler width in cell $i$. The profile is normalized such that $\int \phi_{\nu,i}\,\mathrm{d}\nu=1$.
In practice, we evaluate $H(a_i,u_i)$ using the real part of the complex Faddeeva function, $w(x)$,
\begin{equation}
H(u_i,a_i) = \Re\!\left[w(u_i + \mathrm{i}\,a_i)\right].
\label{eq:H}
\end{equation}
The Doppler width is determined from the Doppler parameter $b_i$ via
\begin{equation}
\Delta \nu_{\mathrm{D},i} = \frac{b_i}{c}\,\nu_0,
\label{eq:dnuD}
\end{equation}
where the total Doppler parameter combines thermal and turbulent broadening,
\begin{equation}
b_i^2 = b_{\mathrm{th},\,i}^2 + b_{\mathrm{turb}}^2.
\label{eq:b_total}
\end{equation}
The thermal contribution is given by
\begin{equation}
b_{\mathrm{th},\,i} = \sqrt{\frac{2k_{\mathrm B}T_i}{A_{\mathrm r}\,m_{\mathrm p}}},
\label{eq:b_th}
\end{equation}
with $T_i$ the gas temperature, $A_{\mathrm{r}}$ the atomic mass, and $m_{\mathrm{p}}$ the proton mass. 

The turbulent term $b_{\mathrm{turb}}$ represents either unresolved small-scale random motions or a specific turbulent feedback model (e.g. the model proposed in Eq.~\ref{eq:b_turb}). The absorption cross section per unit frequency is then
\begin{equation}
\sigma_{\nu,i} =
\frac{\pi e^2}{m_\mathrm{e} c}\,f_{\mathrm{osc}}\,\phi_{\nu,\,i},
\label{eq:sigma_nu}
\end{equation}
where $e$ is the electron charge, $m_\mathrm{e}$ the electron mass, and $f_{\mathrm{osc}}$ the oscillator strength.
The optical depth along a sightline $s$ is obtained by summing the contributions from all intersected cells,
\begin{equation}
\tau_\nu = \sum_{i\in S} n_{\mathrm{ion},\,i}\,\sigma_{\nu,\,i}\,\Delta s_i,
\label{eq:tau}
\end{equation}
where $n_{\mathrm{ion},\,i}$ is the ion number density and $\Delta s_i$ the path length through cell $i$.
Finally, the transmitted flux is given by
\begin{equation}
F_\nu = F_0\,\exp(-\tau_\nu),
\label{eq:intensity}
\end{equation}
with $F_0$ the continuum level. In our implementation, spectra can be computed either on a shared global frequency grid for all transitions or on individual grids centered on each line, depending on the desired application.

\section{Magnesium doublet ratios}
\label{app:mgII_doublet_ratio}

\begin{figure}
    \centering
    \includegraphics[width=\linewidth]{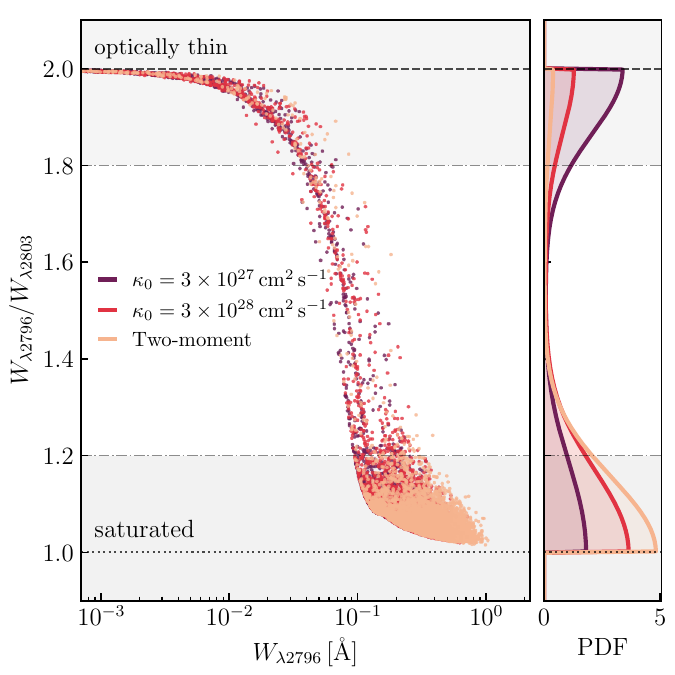}
    \caption{\ion{Mg}{II} doublet ratio as a diagnostic of line saturation. The main panel shows the ratio $\mathcal{R}_{\ion{Mg}{II}}=W_{\lambda2796}/W_{\lambda2803}$ as a function of the rest-frame EW of the stronger \ion{Mg}{II}$\,\lambda2796$ transition for individual sightlines. The right-hand panel shows the corresponding kernel-density estimate of the $\mathcal{R} _{\ion{Mg}{II}}$ distribution. The horizontal dotted and dashed lines mark the saturated and optically thin limits, $\mathcal{R}_{\ion{Mg}{II}}=1$ and $\mathcal{R}_{\ion{Mg}{II}}=2$, respectively. Faster CR transport shifts the \ion{Mg}{II} population toward lower doublet ratios, indicating that an increasing fraction of sightlines becomes saturated.
    }
    \label{fig:mgII_doublet_ratios}
\end{figure}

As an additional diagnostic of line saturation, we analyze the \ion{Mg}{II} doublet ratio \citep[e.g.][]{Nestor2005},
\begin{equation}
    \mathcal{R}_{\ion{Mg}{II}} = \frac{W_{\lambda2796}}{W_{\lambda2803}} .
\end{equation}
In the optically thin limit, this ratio approaches the ratio of the oscillator strengths, $\mathcal{R}_{\ion{Mg}{II}}\simeq 2$ (cf. Table~\ref{tab:ions}), whereas saturated absorption drives the two EWs toward similar values, $\mathcal{R}_{\ion{Mg}{II}}\simeq 1$ \citep[e.g.][]{Nestor2005, Farina2014}. The distribution of $\mathcal{R}_{\ion{Mg}{II}}$ therefore provides a useful consistency check on whether changes in \ion{Mg}{II} EW are mainly driven by increasing column density in optically thin gas or by saturated absorption spread over a broader velocity interval. 

In Fig.~\ref{fig:mgII_doublet_ratios}, we show the \ion{Mg}{II} doublet ratio as a function of $W_{\lambda2796}$ together with the corresponding marginal distribution of $\mathcal{R}_{\ion{Mg}{II}}$. We find a clear dependence on the CR transport model. For slow diffusive transport, $\kappa_0=3\times10^{27}\,\mathrm{cm^2\,s^{-1}}$, most \ion{Mg}{II} absorbers remain close to the optically thin regime. Increasing the diffusion coefficient to $\kappa_0=3\times10^{28}\,\mathrm{cm^2\,s^{-1}}$ shifts the population toward lower doublet ratios, with optically thin and saturated systems contributing in roughly comparable fractions. In the two-moment transport model, the distribution is dominated by $\mathcal{R}_{\ion{Mg}{II}}\simeq 1$, indicating that most \ion{Mg}{II} sightlines are saturated. This trend is consistent with the main results of this work: faster and more self-consistent CR transport promotes the formation of larger columns of cool gas, causing \ion{Mg}{II} absorption to become saturated over an increasing fraction of sightlines.

\section{Geometric contribution to the radial EW trend}
\label{app:path_length_effect}

\begin{figure*}[ht!]
    \centering
    \includegraphics[width=\linewidth]{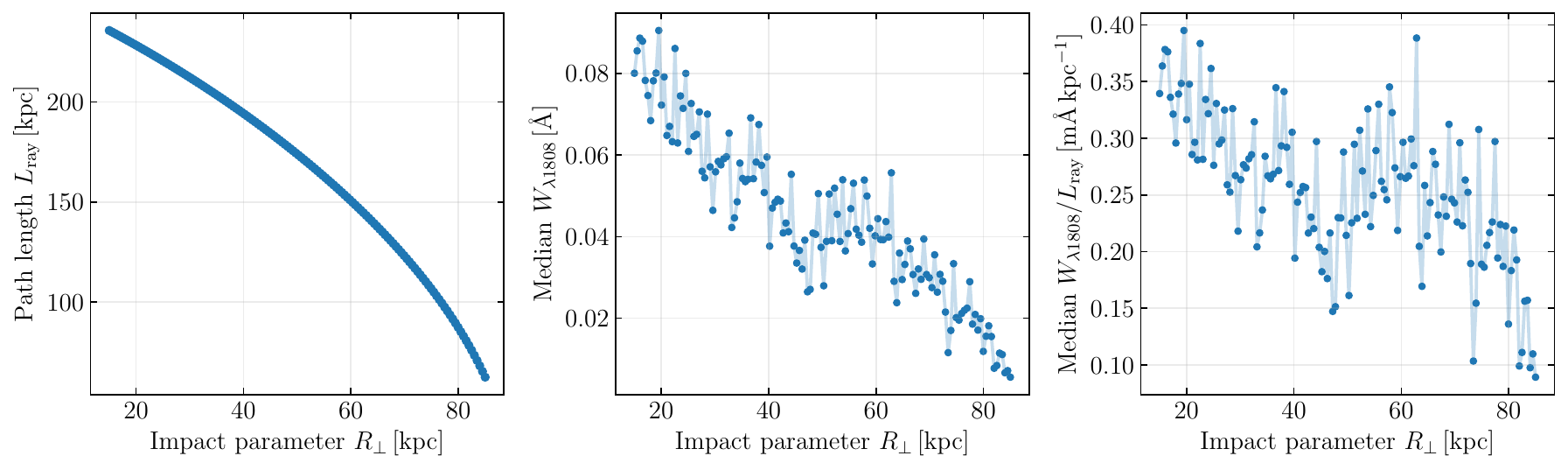}
    \caption{
    Geometric contribution to the radial EW trend for \ion{Si}{II} in the simulation employing two-moment CR transport. Left: effective path length of the curved rays as a function of impact parameter. Middle: median EW as a function of impact parameter. Right: median EW divided by the effective ray path length. The path length decreases with increasing $R_\perp$, contributing to the radial decline of the EW. However, the path-length-normalized EW still decreases with impact parameter, although more weakly, indicating that the radial EW trend also reflects the physical decline of the absorbing gas distribution.
    }
    \label{fig:path_length_effect}
\end{figure*}

Our curved-ray construction is designed to mimic sightlines through a spherical CGM. Consequently, the effective path length through the halo depends on impact parameter. In a spherical geometry, sightlines at larger $R_\perp$ intersect a shorter chord through the
halo. A similar decrease of path length with impact parameter is therefore also expected for realistic galaxy halos. In our analysis, we compute the effective path length of each traced ray from the sequence of intersection points $i$ along the Voronoi mesh,
\begin{equation}
    L_\mathrm{ray} = \sum_i \left| \boldsymbol{x}_{i+1}-\boldsymbol{x}_i \right|
    =
    \sum_i \left[\sum_{j=1}^{3}\left(x_{i+1,j}-x_{i,j}\right)^2\right]^{1/2},
\end{equation}
where the inner sum extends over three coordinates labeled by $j$. This corresponds to the numerical path length used in the ray tracing and includes the actual discretized trajectory through the replicated simulation domain. The resulting geometric variation can contribute to radial trends in integrated quantities such as
column density and EW. 

Figure~\ref{fig:path_length_effect} illustrates this effect for our synthetic \ion{Si}{II} absorption. The left panel shows that the effective ray path length decreases monotonically with impact parameter, as expected from the spherical sightline geometry. The middle panel shows that the median EW follows a similar declining trend. Part of this decrease is therefore geometric, because sightlines at larger $R_\perp$ traverse a shorter path through the halo and sample less absorbing material.

To separate this geometric contribution from changes in the local absorption strength, the right panel shows the EW divided by the effective ray path length. This ratio also decreases with impact parameter, but more weakly than the EW itself. The radial decline in
EW is therefore not caused solely by the decreasing path length. Instead, it reflects a combination of geometry and the physical radial structure of the halo, including the declining gas density, reduced cool-gas CF, and changing ion distribution
at larger radii.

This diagnostic confirms that the curved-ray geometry affects the absolute normalization and radial slope of the EW profiles. However, the same ray construction is applied to all CR transport models. The relative differences between transport models are therefore not driven by this geometric effect, but reflect genuine differences in the amount and kinematic distribution of absorbing gas.

\end{appendix}

\end{document}